\documentclass[12pt]{article}

\usepackage{scicite}

\usepackage{times}

\usepackage{graphicx}
\usepackage[space]{grffile}
\usepackage{latexsym}
\usepackage{amsfonts,amsmath,amssymb}
\usepackage{url}
\usepackage[utf8]{inputenc}
\usepackage{fancyref}
\usepackage{hyperref}
\usepackage{supertabular}
\usepackage{subfigure}
\hypersetup{colorlinks=false,pdfborder={0 0 0},}

\newenvironment{sciabstract}{%
\begin{quote} \bf}
{\end{quote}}

\newcounter{lastnote}

\title{Stellar Activity Masquerading as Planets in the Habitable Zone of the M
dwarf Gliese 581}

\author
{Paul Robertson$^{1,2}$, Suvrath Mahadevan$^{1,2,3}$, Michael Endl$^{4}$, Arpita Roy$^{1,2,3}$\\
\\
\normalsize{$^{1}$Department of Astronomy and Astrophysics, The Pennsylvania State University}\\
\normalsize{$^{2}$Center for Exoplanets \& Habitable Worlds, The Pennsylvania State University}\\
\normalsize{$^{3}$The Penn State Astrobiology Research Center, The Pennsylvania State University}\\
\normalsize{$^{4}$McDonald Observatory, The University of Texas at Austin}\\
}

\date{}

\begin{document} 


\baselineskip24pt


\maketitle


\begin{sciabstract}
The M dwarf Gliese 581 is believed to host four planets, including one
(GJ 581d) near the habitable zone that could possibly support liquid
water on its surface if it is a rocky planet. The detection of another
habitable-zone planet--GJ 581g--is disputed, as its significance depends
on the eccentricity assumed for d. Analyzing stellar activity using the
H$\alpha$ line, we measure a stellar rotation period of $130\pm2$ days
and a correlation for H$\alpha$ modulation with radial velocity.
Correcting for activity greatly diminishes the signal of GJ 581d (to
1.5$\sigma$), while significantly boosting the signals of the other
known super-Earth planets. GJ 581d does not exist, but is an artifact of
stellar activity which, when incompletely corrected, causes the false
detection of planet g.

\end{sciabstract}

At a distance of 6.3 parsecs the M~dwarf Gliese 581 (GJ 581) is believed to host a system of planets discovered using the Doppler radial velocity (RV) technique \cite{bonfils05,udry07,mayor09} and a debris disk \cite{lestrade12}. It is considered a local analog to compact M~dwarf planetary systems found by the \textit{Kepler} spacecraft \cite{dressing13,kopparapu13b}.  

While the periods and orbital parameters of the inner planets b ($P=5.36$ days) and c ($P=12.91$ days) are unchanged since their initial discovery \cite{bonfils05,udry07}, the period of planet d was revised from 82 to 66 days \cite{udry07,mayor09} upon the discovery of a fourth planet e ($P=3.15$ days). Using a combination of data from the HARPS and HIRES spectrographs, planets f ($P=433$ days) and g ($P=36.5$ days) were reported \cite{vogt10}, and their existence promptly questioned \cite{forveille11} using additional data from HARPS. While the reported planet f is now believed to not exist, the existence of planet g is still claimed by some groups. \cite{vogt12} argues for a 5-planet circular model (including g) based on dynamical stability, while studies based on Bayesian statistics \cite{gregory11}, and correlated noise \cite{baluev13,hatzes13b} find no evidence for the existence of g.  At close to half the period of d, its RV signal strongly depends on the eccentricity assumed for the fit to planet d \cite{vogt12,tds12}.  Planet d itself has been questioned; an analysis using a correlated noise model \cite{baluev13} reduced its significance to $2\sigma$, although d is still widely believed by the community. The Gliese 581 system is also of great interest because three of the planets (c,d,g) have all been considered at one time to be among the first exoplanets likely to host habitable environments if they were rocky \cite{udry07,selsis07,vogt10,wordsworth11}, with d still being considered an excellent candidate. This system continues to be studied intensively (e.g. \cite{joiner14}). Since  stellar activity is an important source of noise at the RV amplitudes of the GJ 581 planets, we seek herein to investigate the effects of stellar activity on the RVs of GJ 581 in more detail.

Our analysis of correlation between the RVs and stellar activity indicators focuses on the publicly available spectra from the HARPS spectrograph\footnote{Based on data obtained from the ESO Science Archive Facility under request number 58160}. We adopted the newest published HARPS RVs \cite{forveille11} and measured activity indicators for the H$\alpha$ ($I_{\textrm{H}\alpha}$) \cite{robertson13} and Na I D ($I_\textrm{D}$) \cite{diaz07} lines using the spectra.  Using the cross-correlation functions (CCFs) included with the HARPS spectra, we have calculated the bisector inverse slopes (BIS) as well.  We have excluded one spectrum from our analysis, since it is likely a flare event. Visual inspection of the spectra reveals $I_\textrm{D}$ is contaminated by strong night sky sodium emission lines, which are difficult to correct without sky fibers. We therefore constrained our RV analysis to H$\alpha$ to minimize error.

The RVs of GJ 581 are dominated by the signal of planet b. Upon removing this dominant signal we detect significant anticorrelation in the RV residuals with $I_{\textrm{H}\alpha}$.  For the entire data set (239 spectra), the Pearson correlation coefficient is $r=-0.31$ (probability of no correlation $P(r)=5 \times 10^{-7}$).  Examining the BIS for the HARPS data, we also see evidence for a correlation ($r=0.44$ over the most active period) between BIS and $I_{\textrm{H}\alpha}$, as might be expected if stellar activity is the cause of some RV shifts for GJ 581 (Figure \ref{fig:bis_halpha}).

Frequency analysis of the activity indices indicates that their variability--and thus the induced RV signal--is associated with stellar rotation (Figure \ref{fig:rotps}).  The periodogram of $I_{\textrm{H}\alpha}$ shows a strong peak at 125 days, with an additional peak at 138 days caused by phase changes in the rotation signal. The best fit to the time-series $I_{\textrm{H}\alpha}$ data is obtained when we model individual sinusoids to the periods from December 2005-September 2007 and January 2010-July 2011, when the $I_{\textrm{H}\alpha}$ RMS is twice as high as during the ``quiescent" periods in 2005 and 2009.  The sinusoidal models yield an average rotation period of $130 \pm 2$ days, while the phase and (to a lesser extent) amplitude vary.

The changing phase and amplitude of the activity-induced stellar rotation signal induces a variable effect on the RVs, implying that the slope of the RV-activity correlation is not strictly constant.  Instead of evaluating the RV-activity correlation as one fit over the entire dataset, we have examined RV as a function of $I_{\textrm{H}\alpha}$ over each observing season in the HARPS archive. For the December 2005-September 2007 and January 2010-July 2011 timeframes we have combined two seasons, since $I_{\textrm{H}\alpha}$ shows a coherent rotation signal across the seasons, suggesting the activity behavior remains approximately constant over these times.

\begin{figure}[h!]
\begin{center}
\includegraphics[width=0.7\columnwidth]{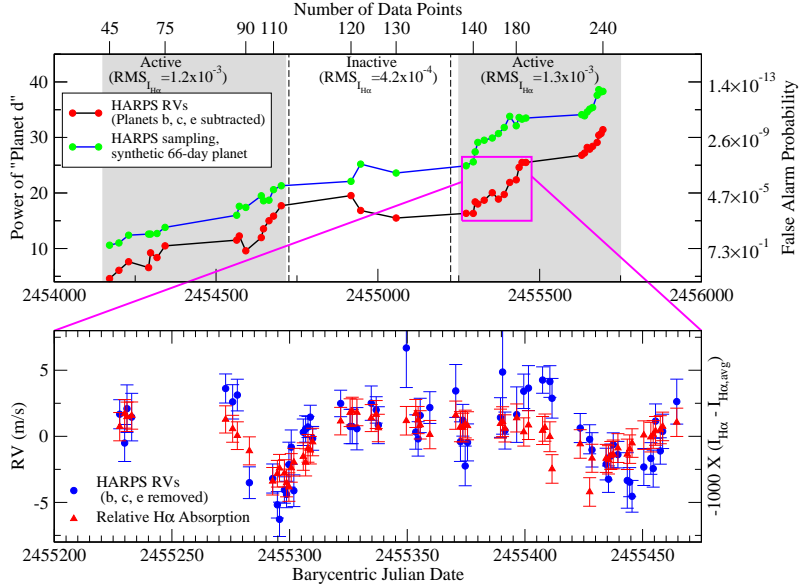}
\caption{\label{fig:harv}
\textit{Top}: Periodogram power of d as a function of time and number of HARPS RVs.  The power is reported at every 5 observations.  We show the actual RVs after removing planets b, c, and e (\textit{black/red}), along with the Keplerian signal of an eccentric 66-day planet sampled with the timestamps and uncertainties of the HARPS sampling (\textit{blue/green}).  Qualitative levels of stellar activity, based on the $I_{\textrm{H}\alpha}$ index, are shown for different time periods.  \textit{Bottom}: HARPS RVs (\textit{blue}) for the region outlined in pink.  Planets b, c, and e have been modeled and removed from the RVs.  We overlay our H$\alpha$ index (\textit{red}), scaled to facilitate visual comparison.  RV and H$\alpha$ are strongly correlated, indicating the remaining Doppler signal is caused by stellar activity.}
\end{center}
\end{figure}

We find significant RV-$I_{\textrm{H}\alpha}$ anticorrelations ($r = -0.45$, $-0.55$, $-0.48$) for three of the five epochs.  These three epochs have RMS values of $1.22 \times 10^{-3}$, $1.10 \times 10^{-3}$ and $1.31 \times 10^{-3}$ in $I_{\textrm{H}\alpha}$, as opposed to $RMS_{I_{\textrm{H}\alpha}} = 4.81 \times 10^{-4}, 4.19 \times 10^{-4}$ in the other seasons, suggesting the star is approximately twice as active during these times.   While we have removed planets b, c, and e from the RVs, we find that once planet b has been removed, the correlation coefficients do not change significantly before/after removing c and e.  The anticorrelation is particularly striking for the 2010 season, shown in Figure \ref{fig:harv}.

\begin{figure}[h!]
\begin{center}
\includegraphics[width=0.7\columnwidth]{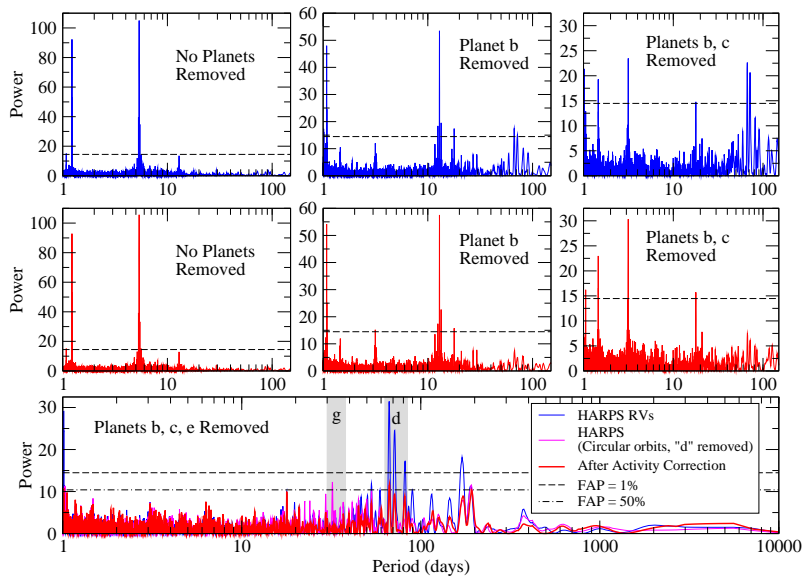}
\caption{\label{fig:pscomp} 
Periodograms for the HARPS RVs before (blue) and after (red) correcting for stellar activity, with the planets successively subtracted.  In the bottom panel we also show (pink) the periodogram after subtracting four circular Keplerian signals, illustrating that the signal interpreted as 581 g \cite{vogt10,vogt12} was caused by fitting a sinusoidal signal to the 581d signal and performing, in essence, an incomplete correction for stellar activity.}
\end{center}
\end{figure}

We correct the HARPS RVs by subtracting the best-fit RV-$I_{\textrm{H}\alpha}$ relation from each of the three epochs for which we observed a significant anticorrelation, leaving the other epochs unchanged. A new RV model then can be used to evaluate the impact of the activity correction on the known exoplanets, and to determine whether any additional signals exist.  We detect the known planets b,c,e using the generalized Lomb-Scargle periodogram \cite{zk09} (Figure \ref{fig:pscomp}).  For planets c and e we observe significant increases in signal power upon correcting for activity.  The power of planet c increases from 53.5 to 57.6, while the power of e increases from 23.5 to 30.3. Formally, false-alarm probability (FAP, from \cite{baluev13}) scales approximately as $e^{-P}\sqrt{P}$ for a Lomb-Scargle periodogram with power $P$, thus the power increase translates to a decrease in FAP by a factor of $60$ for planet c and $800$ for planet e.

In contrast, the power of planet d drops from 31.4 to 11.6 after we apply our activity correction.  As shown in Figure \ref{fig:harv}, the periodogram power increases as a planet would during periods of high activity, but decreases over the epoch with lowest $RMS_{I_{\textrm{H}\alpha}}$.  Planet  d has a reported period of 66 days, which is roughly equal to half the stellar rotation period, suggesting it is a harmonic that loses significance when the rotation signal is removed via decorrelating with $I_{\textrm{H}\alpha}$.

In Table \ref{tab:orbit} we list the parameters of our 3-planet model to the activity-corrected HARPS RVs.  We fit the RVs using the GaussFit \cite{jefferys88} and SYSTEMIC \cite{meschiari09} software packages, finding good agreement between the two.  Our model does not differ much from previous fits to planets b, c, and e \cite{mayor09,forveille11,vogt12}, except that planet e no longer shows any eccentricity after the activity correction.  In the corrected RVs, the 66-day signal appears only at the $1.5\sigma$ significance level in the residual periodogram, and the 33/36-day signal does not appear at all. We conclude the 3-planet solution with activity-induced variability fully explains the observations.

\begin{table}
\begin{tabular}{lccc}
\hline
\textbf{Orbital Parameter} &     \textbf{Planet b} &  \textbf{Planet c} &      \textbf{Planet e} \\ 
Period $P$ (days)   & $5.3686 \pm 0.0001$   & $12.914 \pm 0.002$    & $3.1490 \pm 0.0002$ \\
Periastron Passage $T_0$ & $4751.76 \pm 0.01$ & $4759.2 \pm 0.1$ & $4752.33 \pm 0.05$ \\
(BJD - 2 450 000) & & & \\
RV Amplitude $K$ (m/s) & $12.6 \pm 0.2$ & $3.3 \pm 0.2$ & $1.7 \pm 0.2$ \\
Eccentricity $e$ & $0.00 \pm 0.03$ & $0.00 \pm 0.06$ & $0.00 \pm 0.06$ \\
Semimajor Axis $a$ (AU) & $0.04061 \pm 0.00003$ & $0.0721 \pm 0.0003$ & $0.02815 \pm 0.00006$ \\ 
Minimum Mass $M \sin i$ ($M_{\oplus}$) & $15.8 \pm 0.3$ & $5.5 \pm 0.3$ & $1.7 \pm 0.2$ \\ 
\hline
Zero-point RV offset (m/s) & $-0.52 \pm 0.1$ & & \\
RMS (m/s) & $2.12$ & & \\
\hline & & & \\
\end{tabular}
\caption{Orbital solution for GJ 581 planets after correcting for stellar activity.}
\label{tab:orbit}
\end{table}

We assert the periodic RV signal at 66 days is an artifact induced by the stellar rotation rather than an exoplanet.  Previous studies \cite{mayor09,vogt10} discounted starspot-induced rotational modulation as the origin of RV signals corresponding to planets d and g because the low photometric variability of the star suggests any spots present should be too small to create the observed signals.  However, spatially localized magnetic activity has been observed to influence RV in M stars without producing spots with high optical contrast.  Striking similarities exist with our observed activity-RV correlation results for GJ 581 and those reported for Barnard's star, a slow rotating, photometrically quiet M dwarf of similar spectral type.  Observations with the UVES spectrograph found an RV-$I_{\textrm{H}\alpha}$ anticorrelation  value $r = -0.498$, almost identical  to that observed herein \cite{kurster03}.  K\"urster et al.~attribute this phenomenon to magnetically active regions that stimulate H$\alpha$ emission but do not produce spots of high contrast.  These regions impede local convection, leading to an RV shift.   
Evidence for a relation between chromospheric activity, H$\alpha$ emission, and convective suppression has also been observed in the form of a temperature/radius dependence on H$\alpha$ activity for low-mass stars \cite{stassun12}.    
The anticorrelation (i.e. the negative slope) for RV versus $I_{\textrm{H}\alpha}$ suggests the stellar lines used for RV determination are emitted from a region of convective overshoot.

While the activity-induced RV we observe may not be due to ``typical" dark starspots, localized, rotating regions that magnetically alter the convective velocity field would create RV signatures equivalent to spots. In the absence of simultaneous high-precision photometric monitoring it is difficult to deduce the relative contributions of these different mechanisms. We therefore refer to these as active regions (ARs) hereafter.
A single rotating starspot creates an RV signal at the rotation period, and injects power at a number of its harmonics, primarily $P_{rot}/2, P_{rot}/3,$ and $P_{rot}/4$ \cite{boisse12}. For the purposes of exploring the qualitative impact of such ARs on RV, a starspot model should suffice.  For a rotation period of 130 days, the activity-induced RV signal always includes significant power near the period of planet d. We present two hypotheses for the lack of an observed signal at the periodogram around 130 days. One explanation lies in the geometry of the star and its ARs.  The shape of an activity-induced RV signal changes as a function of the stellar inclination and the latitudes of the ARs, sometimes transferring RV power out of the rotation period into its harmonics. As an illustration, in an analysis using the SOAP code \cite{boisse12} (Figure \ref{fig:rot_lat}), for an inclination of $50^{\circ}$ for the star (consistent with that of its debris disk \cite{lestrade12}) and spots near the stellar equator, the spot-induced RVs are dominated by the 66-day signal. A more important factor, though, is that for the 2010-2011 observing epoch the H$\alpha$ activity contains two signals of roughly equal power at 128 and 69 days, indicating the presence of two ARs instead of one.  In this epoch, we find (Figure \ref{fig:harv_res}) the activity-RV correlation is driven by the 69-day signal rather than the rotation period, indicating activity is injecting RV power at half the rotation period while two ARs are present.  The addition of power at half the rotation period for 109 of 240 observations explains the dominance of the $P_{rot}/2$ signal.

\begin{figure}[h!]
\begin{center}
\includegraphics[width=0.7\columnwidth]{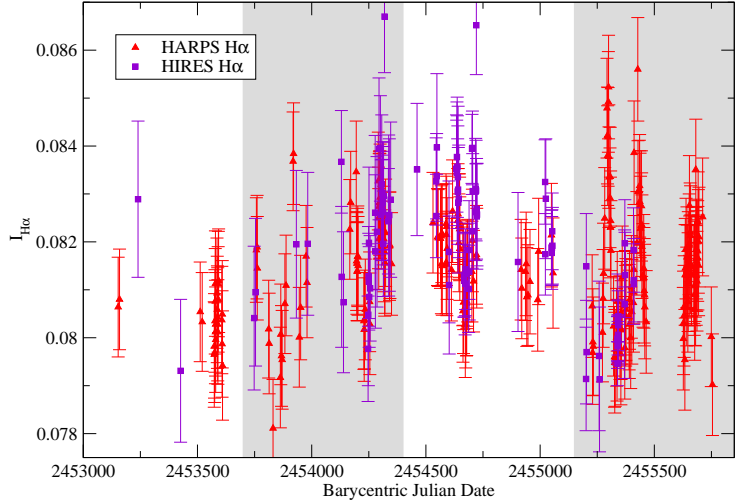}
\caption{\label{fig:hacover}
H$\alpha$ indices from HARPS (\textit{red}) and HIRES (\textit{purple}).  The periods of greatest rotational modulation are shaded.  Note the lack of significant HIRES coverage in the last shaded region, where the signal of the 66-day signal is strongest in both RV and $I_{\textrm{H}\alpha}$.}
\end{center}
\end{figure}

Our activity analysis also helps explain why the signal ascribed to planet d is not observed in the Keck/HIRES RVs alone \cite{baluev13}.  We have computed $I_{\textrm{H}\alpha}$ for the HIRES spectra, which we show alongside the HARPS measurements in Figure \ref{fig:hacover}.  The HIRES data only cover the last portion of the active phase from December 2005-September 2007, and have very little coverage in 2010-2011, where two ARs drive the 66-day period.  Coupled with the higher reported error bars of HIRES compared to HARPS, this yields a non-detection of the 66-day signal.

The  signal of the 33- or 36-day ``planet g", the existence of which has already been called into question, is also an artifact of stellar rotation since no hint of it remains after the activity correction.  Close to $P = P_{rot}/4$, it is another harmonic of the rotation period.  Furthermore, the signal is only observed when fitting a circular orbit to ``planet d," as shown in the bottom panel of Figure \ref{fig:pscomp} \cite{vogt10,forveille11,vogt12}.  By fitting a circular Keplerian model to the 66-day signal, \cite{vogt10} essentially performed an incomplete stellar activity correction, and the signal of ``planet g" was simply leftover noise created by stellar activity.

The impact of stellar activity on the GJ 581 system demonstrates the crucial importance of understanding and treating the presence of activity signals in the quest for low mass planets.  Our activity correction clearly distinguishes between planetary and stellar signals, and reduces the astrophysical noise in the data sufficiently that the signals of very low-mass planets are recovered at much higher significance. This analysis also naturally explains the correlated (red) noise seen \cite{baluev13} in analysis of the HARPS and HIRES data.

Given the advantages of RV surveys of slow-rotating low-mass M dwarfs for RV searches, the physical mechanism that inhibits the convective motion in the stellar atmosphere should be the focus of additional scrutiny. Magnetic activity that impedes the local convective velocity in solar-type stars is invariably accompanied by a dark starspot, so the low contrast of spots in the ARs of Barnard's star and GJ 581 may be a feature unique to low-mass stars. More robust theoretical modeling of magnetohydrodynamics in the atmospheres of old, low-mass stars is required to fully understand this phenomenon. 

GJ 581d and (the now less widely believed) GJ 581g were considered to be among the first exoplanets likely to host habitable environments if they were rocky \cite{vogt10,wordsworth11}.  Given the small number of habitable-zone (HZ)\cite{kopparapu13} planets discovered by Doppler surveys around M~dwarfs, the removal of GJ 581d impacts the RV-based estimate of $\eta_{\oplus}$ (the fraction of stars hosting low-mass planets in their HZs) around M stars. This has been estimated as $\eta_{\oplus} = 0.41^{+0.54}_{-0.13}$ by the HARPS M dwarf survey \cite{bonfils13}.  The exclusion of GJ 581d reduces the rate to 33\%, still within the stated error bars.  More precise estimates of $\eta_{\oplus}$ for M stars from \emph{Kepler} exist (e.g. \cite{dressing13}), but the various HZ limits used by these estimates prevent direct comparison.  While GJ 581 may still be dynamically capable of accommodating terrestrial-mass planets in its HZ, we see no evidence at this time for additional planets in the activity-corrected residuals around our 3-planet model.

 \bibliography{gj581_act_sci.bib}
 
\begin{small}
We acknowledge support from NSF grants AST-1006676,  AST-1126413, AST-1310885, PSARC, and the NASA Astrobiology Institute (NNA09DA76A) in our pursuit of precise RVs in the NIR. This work was supported by funding from the Center for Exoplanets and Habitable Worlds. The Center for Exoplanets and Habitable Worlds is supported by the Pennsylvania State University, the Eberly College of Science, and the Pennsylvania Space Grant Consortium.  M.~E. is supported by NASA through the Origins of Solar Systems Program grant NNX09AB30G and grant AST-1313075 from the NSF.  This research has made use of the Keck Observatory Archive (KOA), which is operated by the W. M. Keck Observatory and the NASA Exoplanet Science Institute (NExScI), under contract with the National Aeronautics and Space Administration. We thank Chad Bender for assistance in studying telluric absorption in the vicinity of H$\alpha$.  The data produced for this study are included in Table \ref{tab:data} of the Supplementary Materials. \\

\noindent \textbf{Supplementary Materials} \newline
Materials and Methods \newline
Figures \ref{fig:rotps}-\ref{fig:bis_halpha} \newline
Tables \ref{tab:hafit}-\ref{tab:data} \newline
References \cite{gds11,pojmanski04,vitale13,engle11,basturk11,wright13} \newline

\end{small}

\clearpage

\setcounter{page}{1}

\makeatletter 
\renewcommand{\thesection}{S\@arabic\c@section}
\makeatother

\section{\textbf{Materials and Methods}}

\subsection{Confirming the Rotation Period}
\label{sec:prot}

\cite{vogt10} claim a rotation period of 94 days, based on $V$-band photometric observations using a Tennessee State University telescope mounted on the roof of Dyer observatory.  Instead, we conclude that the rotation period is 130 days.  Publicly available data support the longer period.  In the HARPS spectra, both $I_{\textrm{H}\alpha}$ and $I_{\textrm{D}}$ show a highly significant signal at $125$ days.  While $I_{\textrm{H}\alpha}$ shows an additional signal near 138 days, we have verified this is the result of a phase offset between the rotation signals observed in the December 2005-September 2007 and January 2010-July 2011 epochs.  $I_{\textrm{D}}$ also shows a second signal at $\approx1300$ days, which is the beat frequency of the 125- and 138-day signals, and therefore likely also related to the phase offset.  This 1300-day period has been previously identified as a stellar activity cycle \cite{gds11,robertson13}, but it may simply be the stellar rotation as observed by the less-frequent cadence used by those surveys.  Alternatively, the cycle may exist, and the two epochs of high activity variability are stimulated by the cycle.  We show our periodograms for $I_{\textrm{H}\alpha}$ and $I_{\textrm{D}}$ in Figure \ref{fig:rotps}.

Fitting a sinusoid to the January 2006-September 2007 H$\alpha$ data, we obtain a rotation period of $132 \pm 4$ days.  For the January 2010-July 2011 observing periods, we have modeled the H$\alpha$ variability as the sum of two sinusoids, finding a period of $128 \pm 4$ days for the stellar rotation.  Assuming each epoch represents a statistically independent sampling of the rotation period, we calculate a final rotation period of $130 \pm 2$ days for GJ 581.

We note that in the 2010-2011 epoch, we observe a highly significant (FAP$=10^{-9}$) signal in the $I_{\textrm{H}\alpha}$ residuals around a single-sinusoid fit.  The periodogram peak lies at 69 days, near half the rotation period.  Fitting a second sinusoid to this period, we find a signal at 69 days with a phase offset of 150 degrees from the rotation signal.  An F-test yields a probability $P=1.2\times10^{-5}$ for the null hypothesis that the data are more consistent with a 1-sinusoid fit, so we conclude the second sinusoid must be statistically significant.  The 2-signal fit with a $\simeq$2:1 period ratio and a phase offset close to 180 degrees is consistent with the presence of two active regions on opposite sides of the star.  We show the periodograms and the 2-signal model for this epoch in Figure \ref{fig:2spot_ps}.  All of our fits to the H$\alpha$ data are listed in Table \ref{tab:hafit}.

Although photometric observations of GJ 581 show little variability, the available data are in fact consistent with a 130-day rotation period.  We include the periodogram of the ASAS $V$-band photometry \cite{pojmanski04} of GJ 581 in Figure \ref{fig:rotps}.  While the signal is not as strong as in the spectral tracers, there is a clear peak at 130 days.  There is also a less significant peak near the purported 94-day period, suggesting that period is an alias of the true signal.  Also, while Vogt et al. did not tabulate their photometric data, their periodogram of the photometry (Figure 1 in their article) shows a broad peak at $125$ days in addition to the 94-day peak.  We therefore conclude that photometric observations of GJ 581 are consistent with our own activity analysis.

Finally, we note that the longer rotation period is more consistent with GJ 581's X-ray luminosity.  SWIFT/XRT observations of GJ 581 yield an X-ray luminosity of $\log(L_X)=26.2$ \cite{vitale13}.  Using an age-rotation-activity curve \cite{engle11}, this luminosity suggests the rotation period should be longer than 100 days.  While this empirical relation by itself is not sufficient to constrain the rotation period, it is another line of evidence in favor of our 130-day period.  Interestingly, the age-rotation-activity curve further indicates GJ 581 should have an age of $\gtrsim10$ Gyr, which would make it the among the oldest known main-sequence stars hosting a multiplanet system.

\subsection{Stellar Activity Correction}

For each observational epoch in which we observe a statistically significant RV-$I_{\textrm{H}\alpha}$ correlation, we have modeled and removed a linear least-squares fit to the data in order to correct the influence of stellar magnetic activity.  The corrections we apply are shown in Figure \ref{fig:harv_epochs} and listed in Table \ref{tab:harv}, along with the associated Pearson correlation coefficients and probabilities of no correlation.  Although we do see evidence for a moderately significant correlation in the 2005 epoch of data, a least-squares fit to the relation yields a slope that is only significant at the $\approx2 \sigma$ level.  Considering our other fits have slopes significant to $4\sigma$ or greater, we elected not to add such an uncertain additional correction to the data.

\subsubsection{Activity-RV Correlation With 2 Active Regions}
\label{sec:2rot}

As mentioned above, the 2010-2011 observing epoch includes compelling evidence for the presence of two active regions on opposite sides of the star.  Because the period of the H$\alpha$ signal induced by the second active region is so similar to the 66-day RV signal, we examined the residual RV-activity correlation for each of the two H$\alpha$ signals present during this time.  

In Figure \ref{fig:harv_res}, we show RV as a function of $I_{\textrm{H}\alpha,res}$, the residual $I_{\textrm{H}\alpha}$ after subtracting one of the two sinusoidal signals listed in Table \ref{tab:hafit}.  The Pearson correlation coefficient for the data with the 128-day rotation signal removed is $r = -0.48$, consistent with the strong anticorrelations observed in other epochs, while $r = -0.16$ with the 69-day H$\alpha$ signal subtracted, indicating no significant correlation.  This suggests--as expected--the RV-activity anticorrelation is driven entirely by the 69-day H$\alpha$ signal during the epoch containing two active regions.  The dominance of the stellar rotation's first harmonic over the fundamental in RV can be better understood in this context, since 109 of the 240 observations come from this epoch.  In order to present the most physically-motivated activity correction possible, we list the activity-RV fit with the 128-day rotation signal removed from $I_{\textrm{H}\alpha}$ in Table \ref{tab:harv}.

It is important to note that although the correction listed in Table \ref{tab:harv} is our best model to the stellar rotation signal, the qualitative results of our analysis are insensitive to how we divide the data into subsets.  Indeed, a significant RV-$I_{\textrm{H}\alpha}$ correlation is present even when considering all the data together.  In Figure \ref{fig:harv_all}, we show RV as a function of $I_{\textrm{H}\alpha}$ for the entire data set after removing the three planets.  If we remove the correlation observed in the overall data set, the power of the 66-day signal is still significantly reduced, but not completely eliminated.

In both our activity and RV analyses, we have excluded the HARPS spectrum taken at $BJD = 2454610.74293237$.  This spectrum has anomalously high emission in both H$\alpha$ and sodium, leading us to conclude it might be associated with a flare event.  We note that this was not one of the spectra omitted from the most recent HARPS analysis \cite{vogt12}.

Upon applying the activity correction, we obtain the orbital model presented in Table \ref{tab:orbit}.  A schematic of this model is shown in Figure \ref{fig:orbit}.

\subsection{Impact of Active Regions on RV}

In order to estimate the impact of active regions on the RVs of GJ 581, we have simulated starspots--which should have similar qualitative RV imprints as active regions--using the Spot Oscillation and Planet (SOAP) code \cite{boisse12}.  As discussed in the article, we have fixed the stellar inclination at $50^{\circ}$, loosely based on the presence of a debris disk at that inclination.  For each simulation, we have sampled the resulting RV signal at the cadence of the HARPS data and retained the HARPS RV uncertainties.

Figure \ref{fig:rot_lat} shows the resulting periodograms from simulated spots at three different latitudes.  We see that for spots closer to the stellar equator, the RV signal shifts from the stellar rotation period to the first harmonic, in agreement with the 66-day signal observed in the real RVs.

For the 2010-2011 observing epochs, we have also simulated the effects of two active regions on the star.  As for the single-AR simulations, we have fixed the stellar inclination at $50^{\circ}$ and sampled the simulated RV signal at the cadence of the HARPS observations during the last two observing seasons.  The ARs are separated in longitude by $180^{\circ}$.  As shown in the periodogram of the simulated RVs (Figure \ref{fig:soap_late}), the signal of ``planet d''--including its aliases at 51 and 88 days--is clearly reproduced by two ARs.  Additionally, while SOAP does not simulate H$\alpha$ emission, its simulated photometry is consistent with our observed activity behavior during this epoch; the periodogram of the photometry again shows peaks at the rotation period and $P_{rot}/2$\footnote{In order for the photometry to remain consistent with our observations, one of the two active regions must be brighter than the other.  This is understandable in the astrophysical context, as any active regions observed in H$\alpha$ will have varying contributions from spots, plage, and filaments.}.

\subsection{Bisector Analysis}

To confirm the signature of stellar activity in the line profiles, we examine the bisectors of the cross-correlation function (CCF) provided by HARPS \cite{basturk11}. Specifically, we use orders in the wavelength range 5432 - 6212 \AA, since bluer orders have lower flux, and redder orders are generally expected to exhibit less dependence on stellar activity (Fig. \ref{fig:s2n}a).

We sum the CCFs (provided by the HARPS reduction) of the selected orders to obtain an aggregate and calculate the bisector inverse slope (BIS) as described in \cite{wright13}. We prefer to calculate the BIS instead of using the CCF bisector span (CBS) measurement provided with the HARPS data in order to customize the order selection. Six spectra ($BJD =$ 2455349.63633994, 2455370.57817460, 2455390.54432217, 2455396.49708365, 2455443.49985669, \newline 2455663.75875103) with low average CCF counts are excluded, since these cause high BIS errors (Fig. \ref{fig:s2n}b). Note that the BIS errors were determined  from the dispersion in each value between different echelle orders. Of the three observing seasons with significant RV-$I_{\textrm{H}\alpha}$ anticorrelation, the third season of interest (January 2010 to July 2011) shows the strongest BIS-$I_{\textrm{H}\alpha}$ correspondence, with a Pearson correlation coefficient $r = 0.36$\footnote{The BIS-$I_{\textrm{H}\alpha}$ is positive, which we expect, since BIS and RV are anticorrelated, and we see RV is anticorrelated with $I_{\textrm{H}\alpha}$.}, in agreement with the previous assertion that this is the temporal period of highest observed activity (Fig. \ref{fig:bis_halpha}).  While the correlation coefficient indicates a high confidence level ($P(r)=1.8\times10^{-4}$), we emphasize that this result is sensitive to the wavelength range (orders) selected for analysis. The variation in bisector shape, as evidenced by the BIS, points to changes in the line profile that affect the RV measurement. This is consistent with the effect of stellar activity.

\setcounter{figure}{0}

\begin{figure}[h!]
\makeatletter 
\renewcommand{\thefigure}{S\@arabic\c@figure}
\makeatother
\begin{center}
\includegraphics[width=0.9\columnwidth]{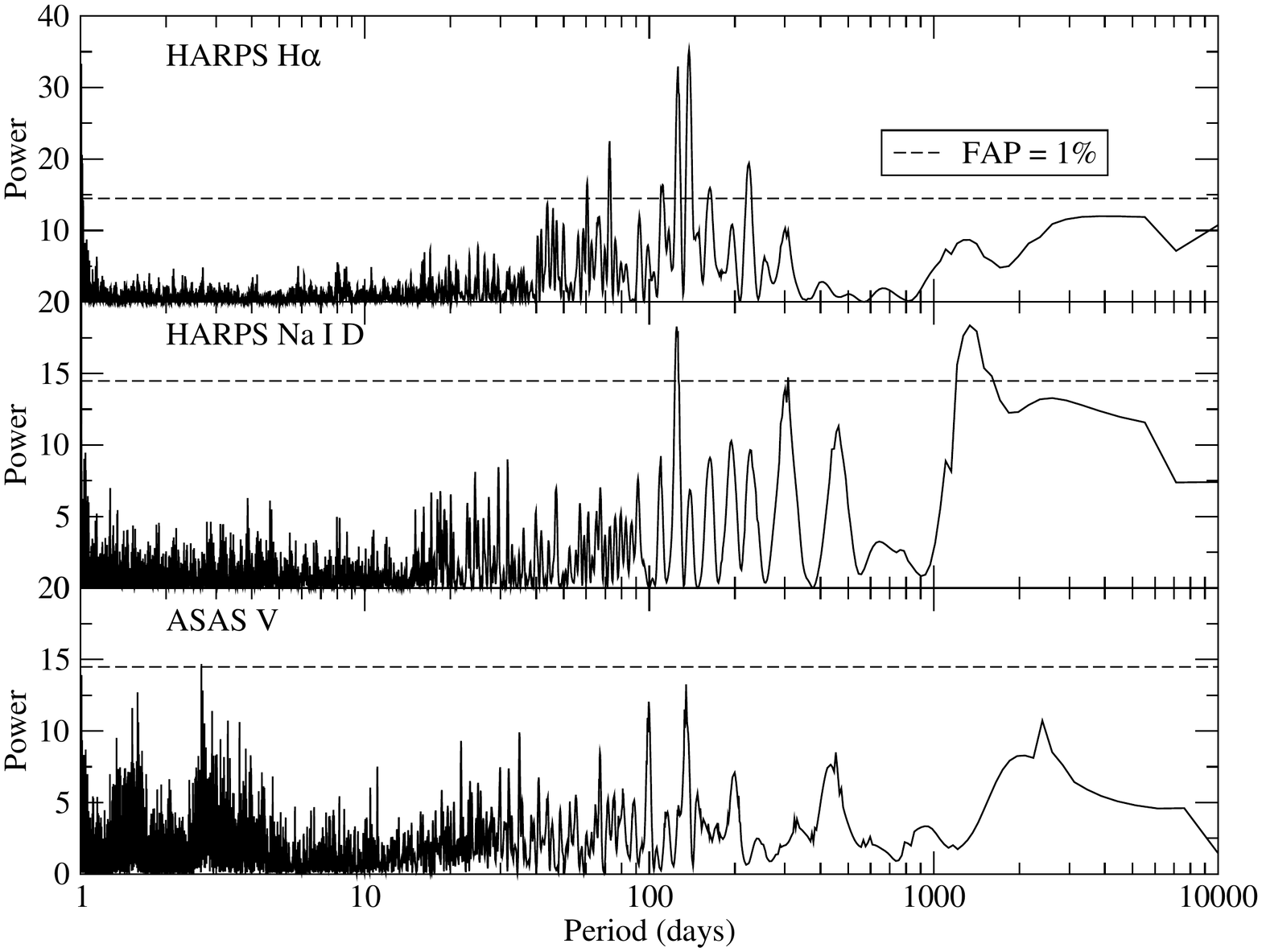}
\caption{\label{fig:rotps}
Generalized Lomb-Scargle periodograms for spectral and photometric activity tracers.  A peak near 130 days appears in all tracers, leading us to conclude it must be the rotation period.  The second peak at 138 days in the H$\alpha$ periodogram is caused by a phase shift in the rotation signal, and the 1300-day peak in the Na I D line is the beat frequency of the two highest peaks in H$\alpha$. The Na I data are noisier than H$\alpha$ due to the presence of telluric emission lines in that spectral region.
}
\end{center}
\end{figure}

\begin{figure}[h!]
\makeatletter 
\renewcommand{\thefigure}{S\@arabic\c@figure}
\makeatother
\begin{center}
\includegraphics[width=0.9\columnwidth]{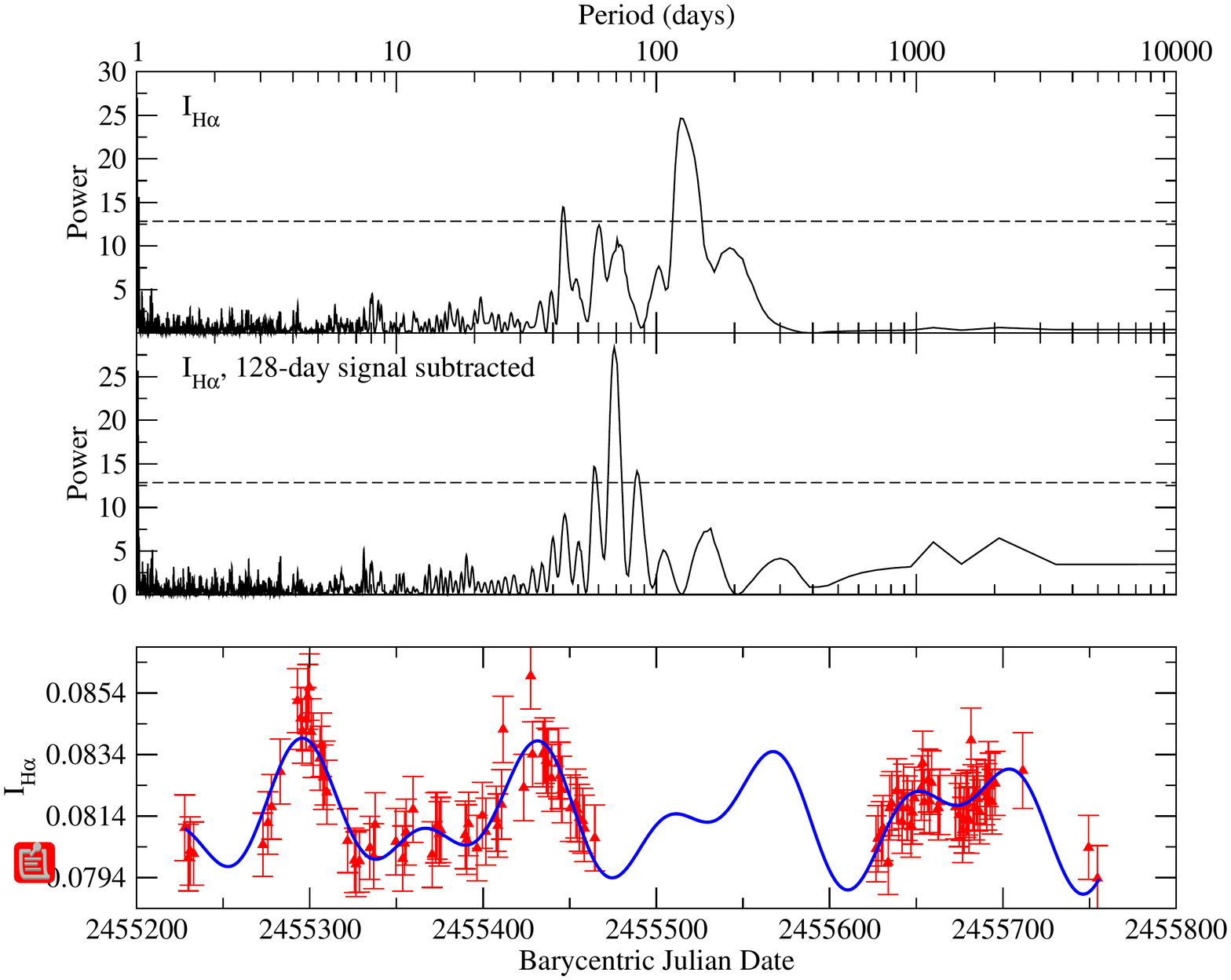}
\caption{\label{fig:2spot_ps}
\emph{Top}: Periodogram of $I_{\textrm{H}\alpha}$ from January 2010-July 2011.  \emph{Middle}: Residual periodogram of $I_{\textrm{H}\alpha}$ of the same data after modeling and removing the 128-day stellar rotation signal.  The dashed lines indicate the power level required for a false alarm probability of 1\% according to \cite{baluev13}.  \emph{Bottom}: $I_{\textrm{H}\alpha}$ from the January 2010-July 2011 observing seasons.  Our two-signal ($P_1 = 128$d, $P_2 = 69$d) model to the data is shown in blue.}
\end{center}
\end{figure}

\begin{figure}[h!]
\makeatletter 
\renewcommand{\thefigure}{S\@arabic\c@figure}
\makeatother
\begin{center}
\includegraphics[width=0.9\columnwidth]{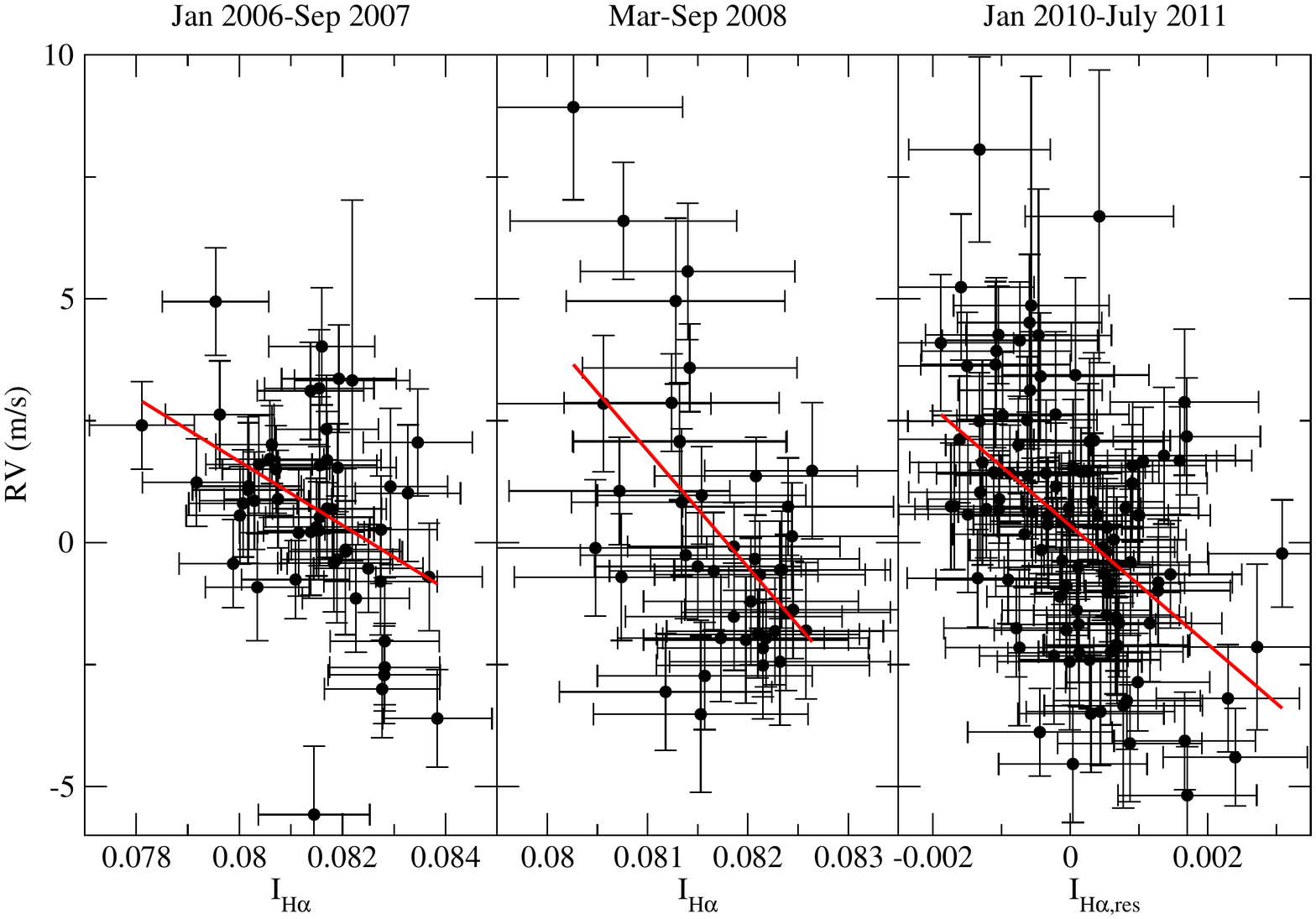}
\caption{\label{fig:harv_epochs}
RV as a function of $I_{\textrm{H}\alpha}$ for each of the three epochs in Table \ref{tab:harv}.  The linear least-squares fit to each epoch is shown as a solid red line.  Note that in the 2010-2011 segment we have removed the 128-day stellar rotation signal from $I_{\textrm{H}\alpha}$ (see Section \ref{sec:2rot}).}
\end{center}
\end{figure}

\begin{figure}[h!]
\makeatletter 
\renewcommand{\thefigure}{S\@arabic\c@figure}
\makeatother
\begin{center}
\includegraphics[width=0.9\columnwidth]{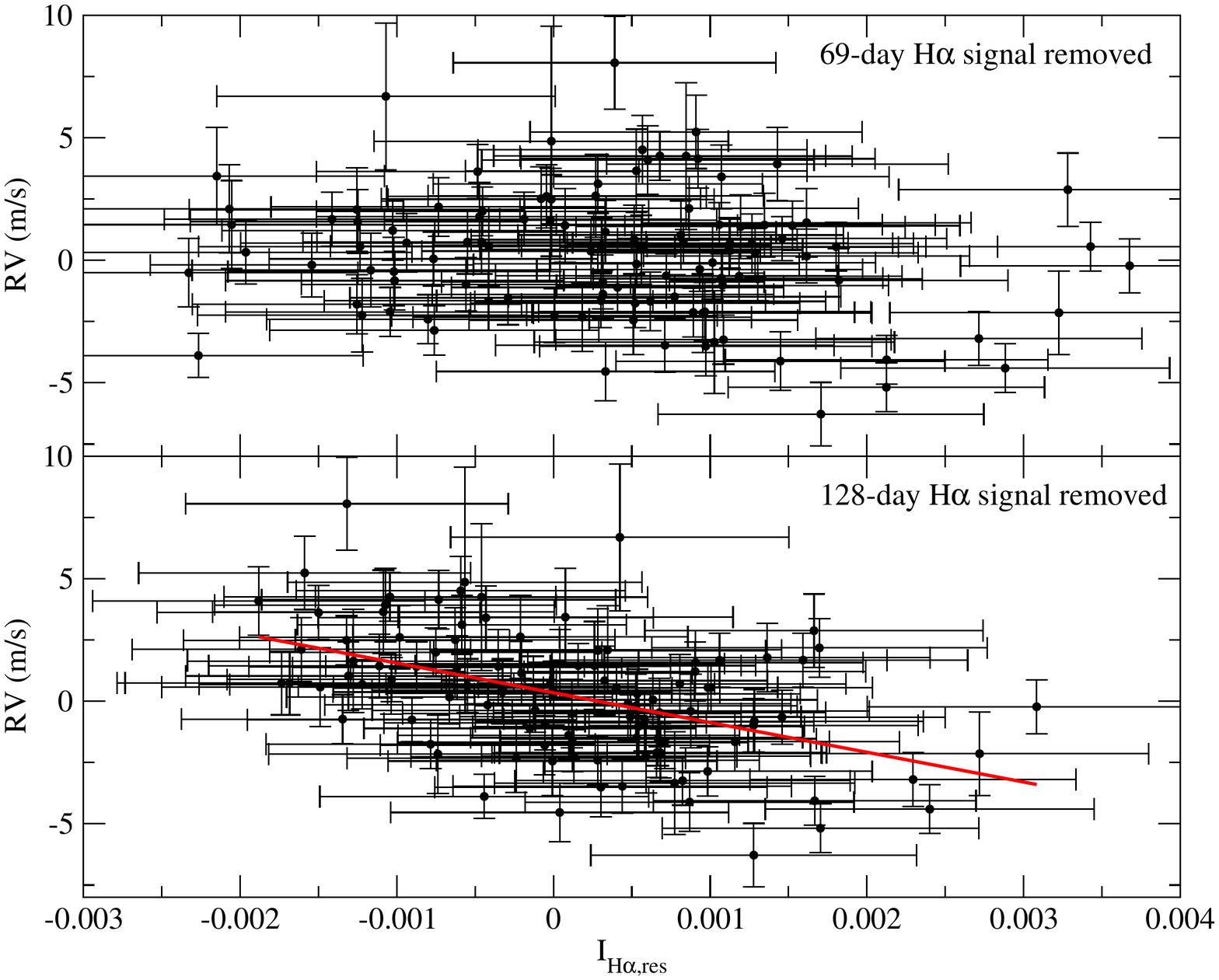}
\caption{\label{fig:harv_res}
RV as a function of $I_{\textrm{H}\alpha}$ for the January 2010-July 2011 observing seasons.  In each panel, one of the two activity signals described in Section \ref{sec:prot} has been removed from $I_{\textrm{H}\alpha}$.  Our linear least-squares fit to the data after subtracting the 128-day rotation signal is shown in red.}
\end{center}
\end{figure}

\begin{figure}[h!]
\makeatletter 
\renewcommand{\thefigure}{S\@arabic\c@figure}
\makeatother
\begin{center}
\includegraphics[width=0.9\columnwidth]{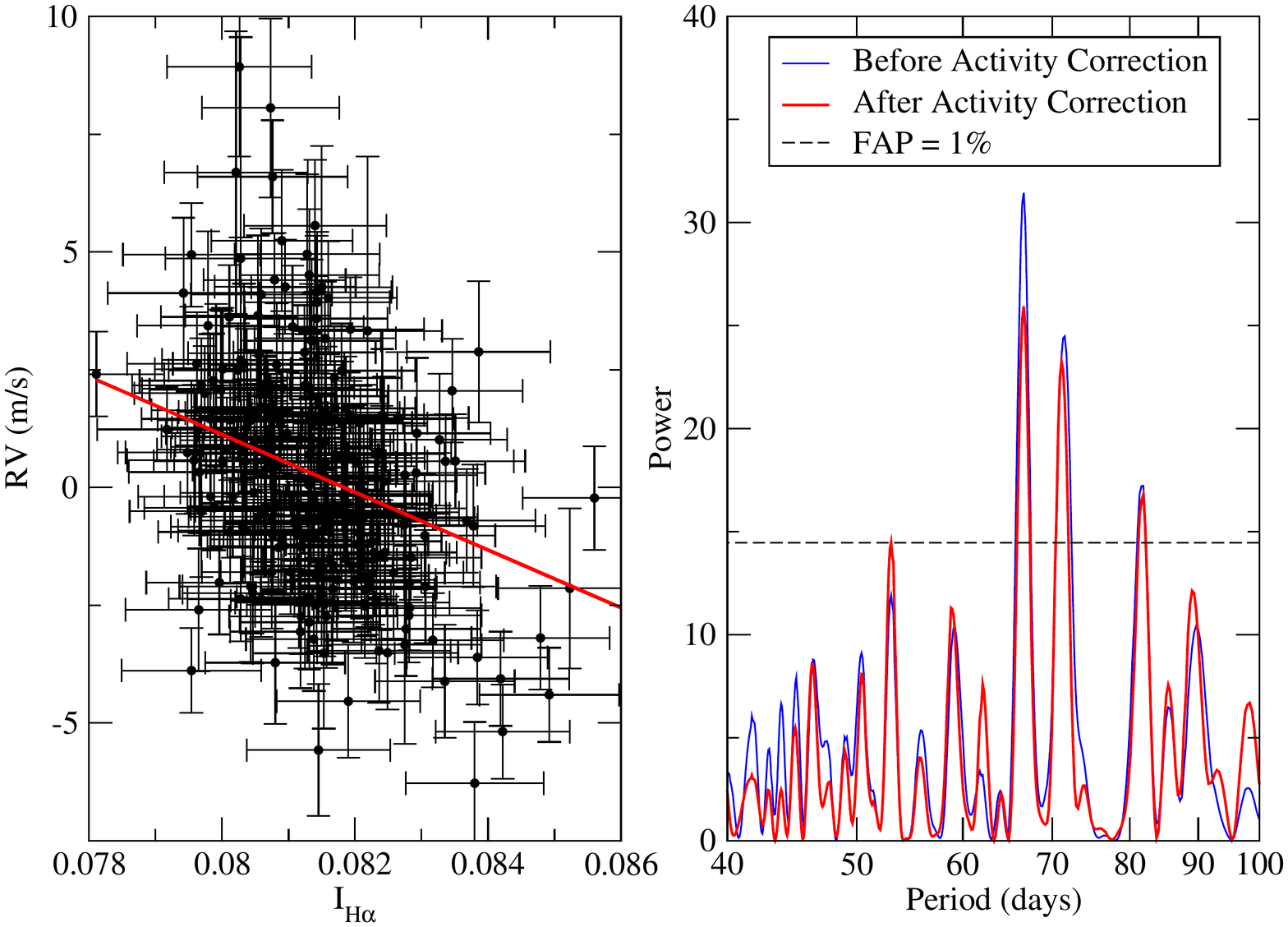}
\caption{\label{fig:harv_all}
\emph{Left}: RV as a function of $I_{\textrm{H}\alpha}$ using all HARPS RVs simultaneously in a single correlation.  Planets b, c, and e have been removed.  Our linear least-squares fit to the data is shown as a solid red line.  \emph{Right}: Generalized Lomb-Scargle periodogram of the HARPS RVs after subtracting planets b, c, and e.  Even when treating the RV-activity dependence over the entire data set, the signal of planet d is significantly reduced.}
\end{center}
\end{figure}

\begin{figure}[h!]
\makeatletter 
\renewcommand{\thefigure}{S\@arabic\c@figure}
\makeatother
\begin{center}
\includegraphics[width=0.9\columnwidth]{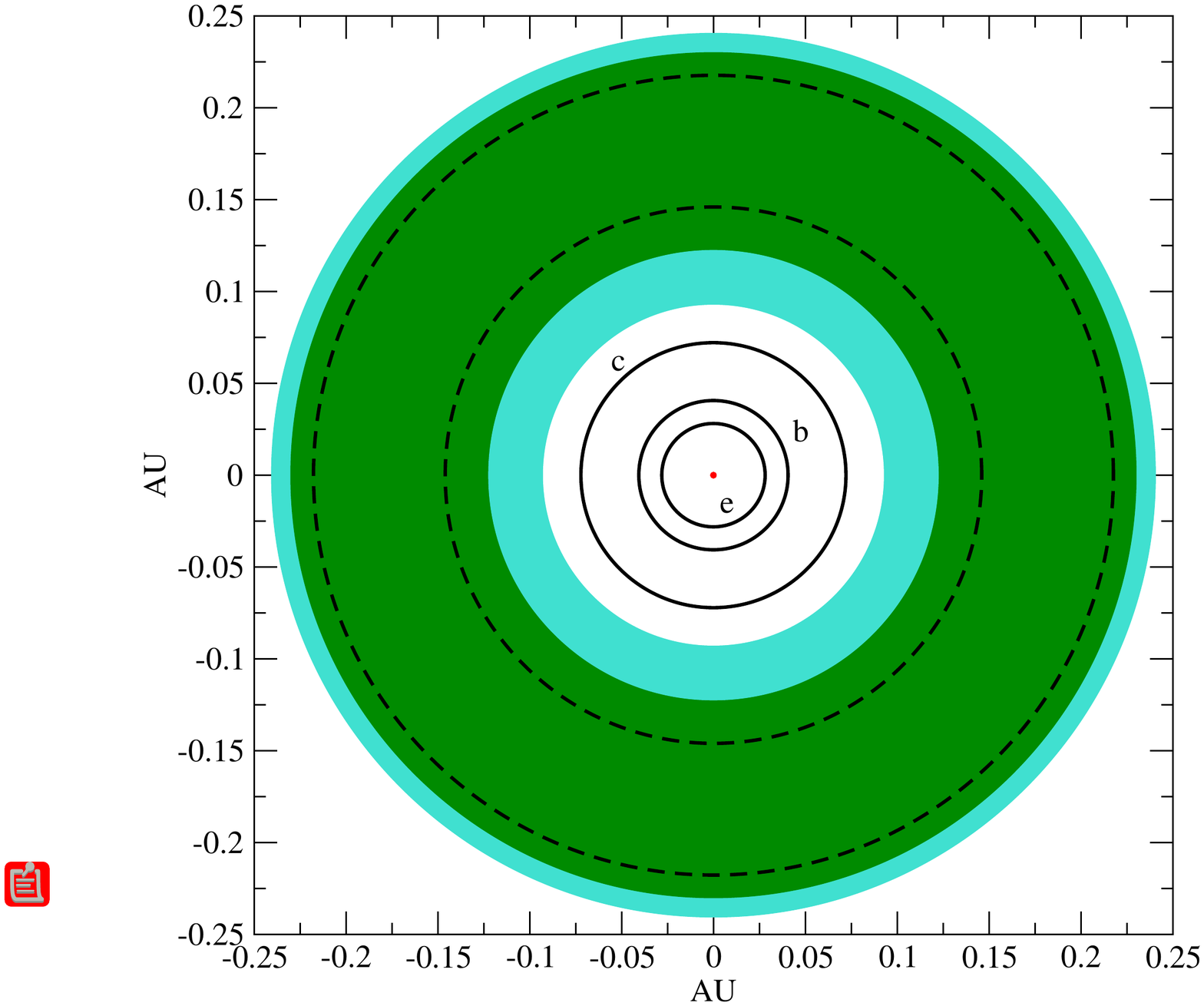}
\caption{\label{fig:orbit}
Schematic of our 3-planet orbital model.  The stellar radius is drawn to scale.  The orbits ascribed to the 36- and 66-day stellar rotation harmonics by \cite{vogt10} are shown as dashed lines, along with the conservative (\textit{dark green}) and optimistic (\textit{light green}) habitable zone boundaries according to \cite{kopparapu13}.}
\end{center}
\end{figure}

\begin{figure}[h!]
\makeatletter 
\renewcommand{\thefigure}{S\@arabic\c@figure}
\makeatother
\begin{center}
\includegraphics[width=0.9\columnwidth]{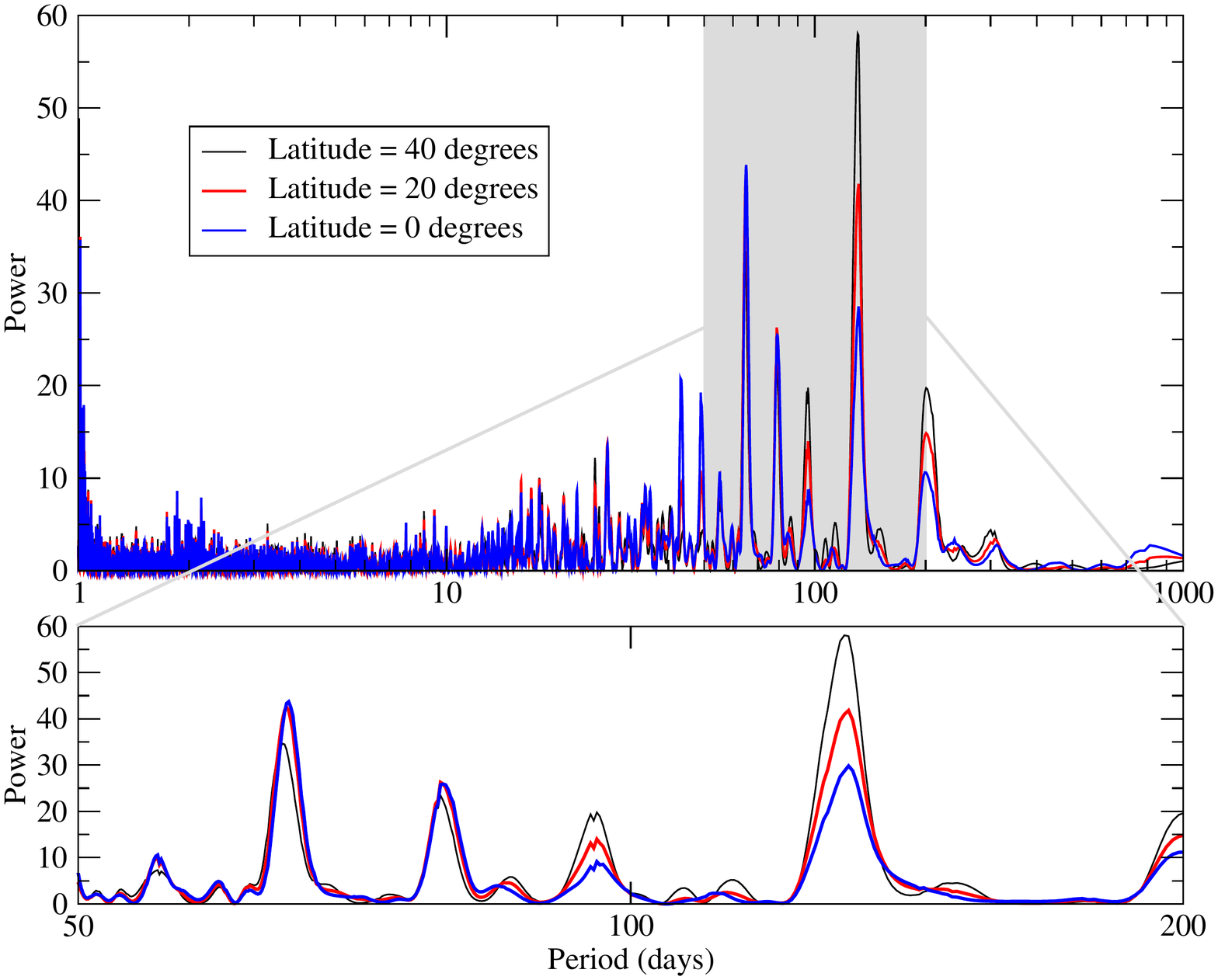}
\caption{\label{fig:rot_lat}
Periodograms of activity-induced RV signals using SOAP \cite{boisse12} for active regions at various latitudes.  The RVs are sampled using the HARPS sampling and errors.  The stellar inclination is fixed at $50^{\circ}$ for this illustration.  As the active regions move closer to the equator, the peak at the rotation period $P_{rot}$ decreases, while the peak at $P_{rot}/2$ increases. We empahsize that this an illustrative model to show how harmonics of the rotation period are created in the RVs. }
\end{center}
\end{figure}

\begin{figure}[h!]
\makeatletter 
\renewcommand{\thefigure}{S\@arabic\c@figure}
\makeatother
\begin{center}
\includegraphics[width=0.9\columnwidth]{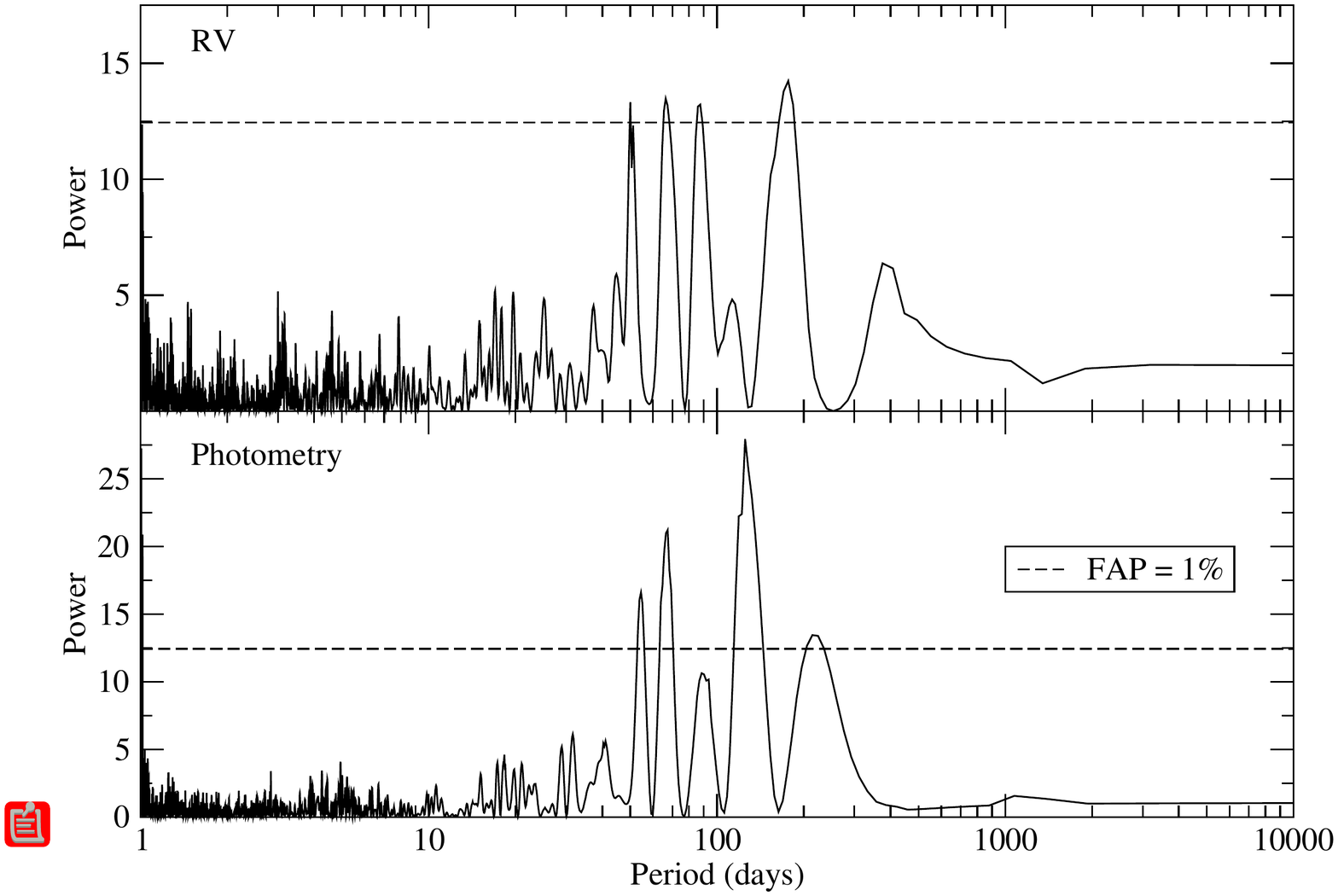}
\caption{\label{fig:soap_late}
Periodograms of activity-induced RV (\emph{top}) and photometric variability (\emph{bottom}) based on SOAP simulations of two active regions during the 2010-2011 epoch.  The rotation period is fixed at 130 days, the stellar inclination is $50^{\circ}$, and the active regions are separated by $180^{\circ}$ in longitude.  The broad peak at 180 days is a 6-month alias, caused by the observing cadence.  The other three highest peaks in the RV periodogram correspond to the period of ``planet d" and its aliases.  For comparison to real data, a periodogram showing the real 66-day RV signal is shown in the bottom panel of Figure \ref{fig:pscomp}, and periodograms of the activity indices is given in Figure \ref{fig:rotps}.
}
\end{center}
\end{figure}

\begin{figure}[h!]
\begin{center}
\makeatletter 
\renewcommand{\thefigure}{S\@arabic\c@figure}
\makeatother
\subfigure[\label{fig:s2n_order}]{\includegraphics[width=0.48\columnwidth]{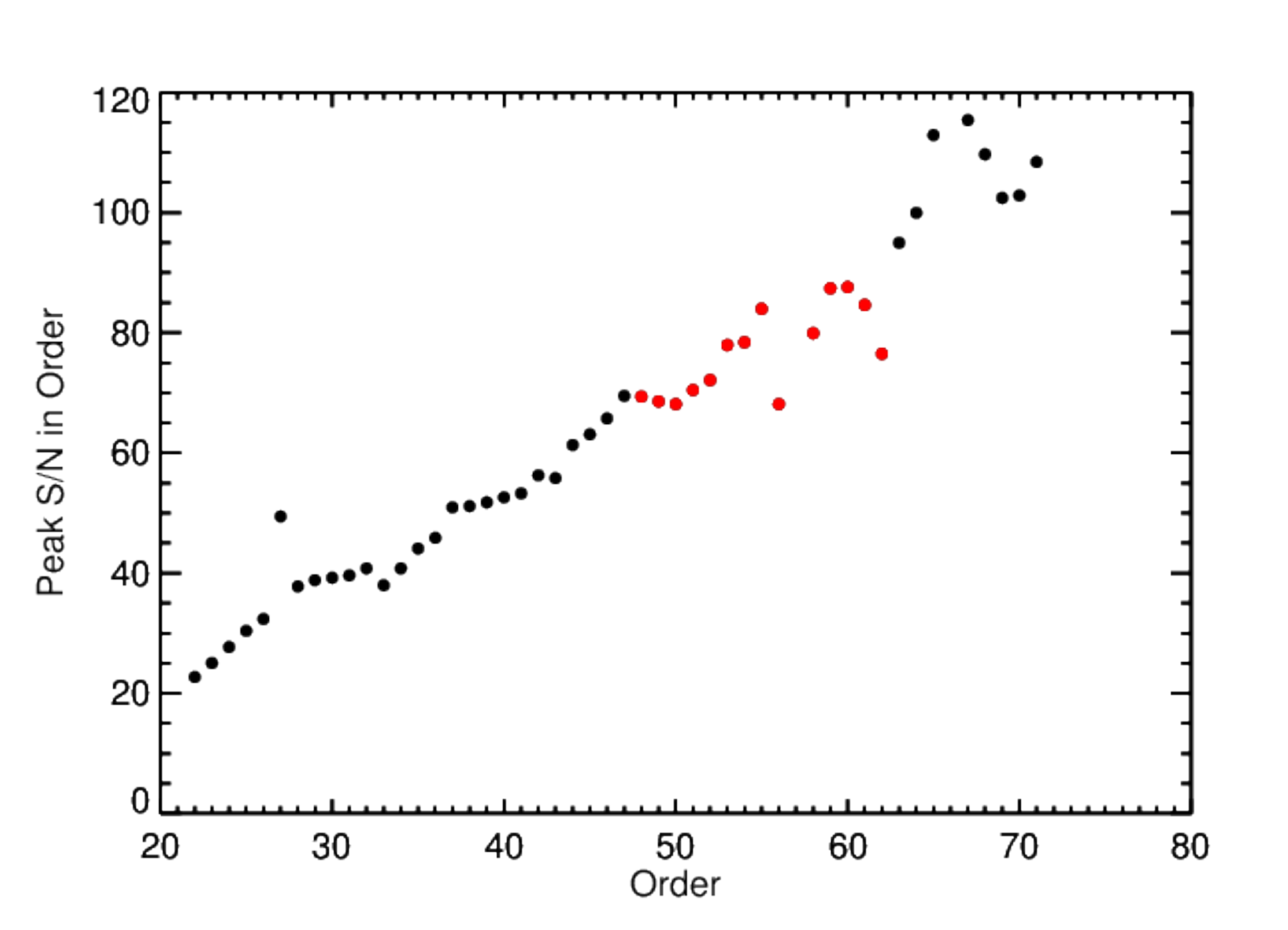}}
\subfigure[\label{fig:s2n_file}]{\includegraphics[width=0.48\columnwidth]{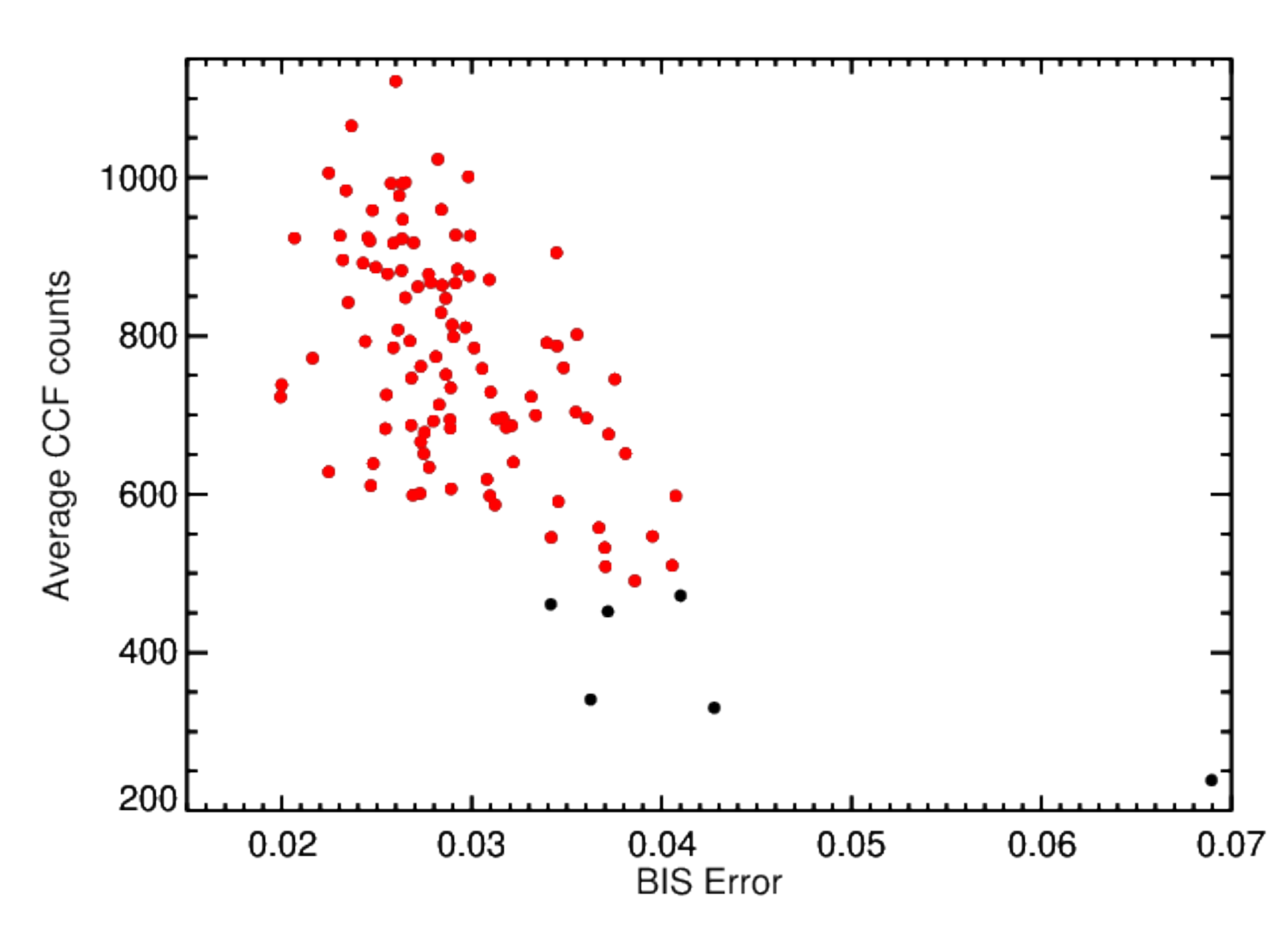}}
\caption{\label{fig:s2n}
\emph{a}: Peak  signal to noise ratio (SNR) in each order, as a method for selecting a continuous set of orders with relatively high flux and sensitivity to activity. Red points show the selected orders
\emph{b}: Average SNR in CCF per velocity bin (161 bins of 0.25 km s$^{-1}$ width), as a threshold for removing spectra with low flux, and thus larger BIS errors. Red points show the selected spectra.}
\end{center}
\end{figure}

\begin{figure}[h!]
\makeatletter 
\renewcommand{\thefigure}{S\@arabic\c@figure}
\makeatother
\begin{center}
\includegraphics[width=0.7\columnwidth]{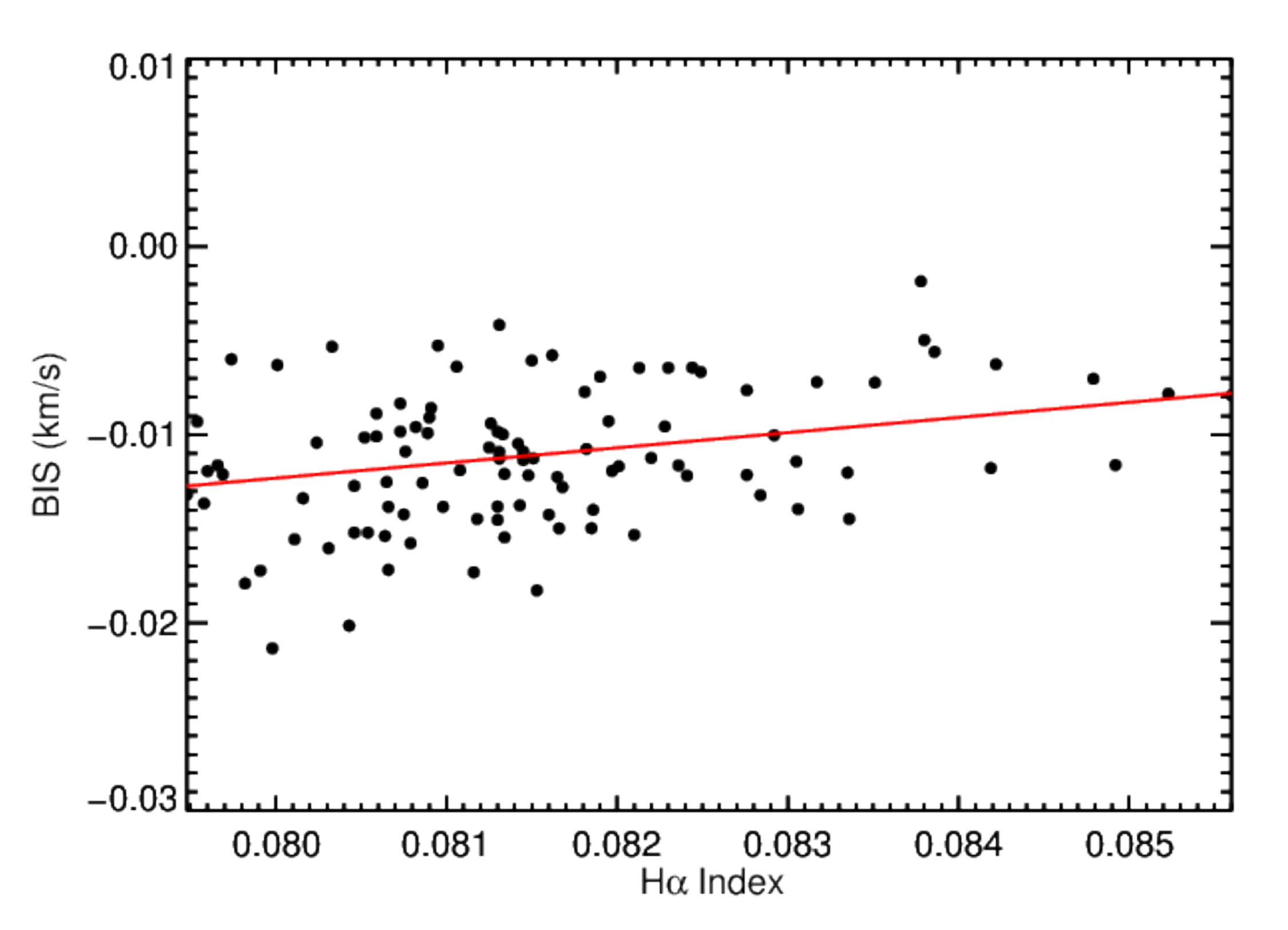}
\caption{\label{fig:bis_halpha}
Correlation between the bisector inverse slope (BIS) and the H$\alpha$ activity index ($I_{\textrm{H}\alpha}$) during the third season of interest (January 2010 to July 2011), with $r = 0.36$.}
\end{center}
\end{figure}

\clearpage

\setcounter{table}{0}

\begin{table}[h!]
\makeatletter 
\renewcommand{\thetable}{S\@arabic\c@table}
\makeatother
\begin{tabular}{lccc}
\textbf{Signal} & \textbf{Period} & \textbf{Amplitude} & \textbf{$T_{\textrm{max}}$} \\
 & (days) & (in $I_{\textrm{H}\alpha} \times 1000$) & (BJD-2450000) \\ \hline
\emph{Jaunary 2006-September 2007} & & & \\
Stellar Rotation & $132 \pm 4$ & $1.4 \pm 0.2$ & $3649 \pm 4$ \\ \hline
\emph{January 2010-July 2011} & & & \\
Stellar Rotation & $128 \pm 4$ & $1.5 \pm 0.1$ & $5168 \pm 3$ \\
Stellar ``Half-Rotation" & $69 \pm 1$ & $1.1 \pm 0.1$ & $5226 \pm 2$ \\
\end{tabular}
\caption{Sinusoidal models to $I_{\textrm{H}\alpha}$ during the epochs for which we observe a stellar rotation signal.}
\label{tab:hafit}
\end{table}

\begin{table}[h!]
\makeatletter 
\renewcommand{\thetable}{S\@arabic\c@table}
\makeatother
\begin{tabular}{lccc}
\textbf{Epoch} & \textbf{RV($I_{\textrm{H}\alpha}$) (m/s)} & \textbf{$r$} & \textbf{$P(r)$} \\
January 2006-September 2007 & $RV = 54(15) - 650(190) \times I_{\textrm{H}\alpha}$ & $-0.45$ & $5 \times 10^{-4}$ \\
March-September 2008 & $RV = 200(50) - 2400(600) \times I_{\textrm{H}\alpha}$ & $-0.55$ & $8 \times 10^{-5}$ \\
January 2010-May 2011 & $RV = 0.35(0.2) - 1200(200) \times I_{\textrm{H}\alpha,res}$ & $-0.48$ & $6 \times 10^{-8}$ \\
\end{tabular}
\caption{Linear least-squares fits to the RV-$I_{\textrm{H}\alpha}$ relation for each of the three observational epochs in which we apply a stellar activity correction.  The RV signals of planets b, c, and e were removed before computing the fits.  For each epoch, we include the Pearson correlation coefficient $r$ and the corresponding probability of no correlation $P(r)$.}
\label{tab:harv}
\end{table}

\clearpage
\makeatletter 
\renewcommand{\thetable}{S\@arabic\c@table}
\makeatother
\begin{center}
\footnotesize
\tablecaption{H$\alpha$ and Na I D activity indices from the publicly-available HARPS spectra, along with their associated RVs from \cite{forveille11}.  We have determined the spectrum at 2454610.74293237 to be anomalously high in both activity indices--potentially indicative of a flare--and have therefore excluded it from our analysis.  The last three spectra have no published RVs, and are therefore excluded from our RV analysis.}
\label{tab:data}
\tablefirsthead{\textbf{BJD - 2 450 000} & \textbf{$I_{\textrm{H}\alpha}$} & \textbf{$\sigma_{I_{\textrm{H}\alpha}}$} & \textbf{$I_{\textrm{D}}$} & \textbf{$\sigma_{I_{\textrm{D}}}$} & RV (m/s) & $\sigma_{RV}$ (m/s) \\ 
 & & & & & \multicolumn{2}{c}{(from \cite{forveille11})} \\ }
 \tablehead{\multicolumn{4}{l}{\textit{Table \ref{tab:data} cont'd}} \\
 \textbf{BJD - 2 450 000} & \textbf{$I_{\textrm{H}\alpha}$} & \textbf{$\sigma_{I_{\textrm{H}\alpha}}$} & \textbf{$I_{\textrm{D}}$} & \textbf{$\sigma_{I_{\textrm{D}}}$} & RV (m/s) & $\sigma_{RV}$ (m/s) \\ }
 
\begin{supertabular}{lllllrl}
3152.71289435 	 & $ 0.08064 $ 	 & $ 0.00104 $ 	 & $ 0.06131 $ 	 & $ 0.00199 $ 	 & $ -10.25 $ 	 & $ 1.10 $ \\
3158.66345975 	 & $ 0.08080 $ 	 & $ 0.00105 $ 	 & $ 0.06627 $ 	 & $ 0.00198 $ 	 & $ -19.05 $ 	 & $ 1.30 $ \\
3511.77334131 	 & $ 0.08054 $ 	 & $ 0.00104 $ 	 & $ 0.06923 $ 	 & $ 0.00186 $ 	 & $ -7.25 $ 	 & $ 1.20 $ \\
3520.74474617 	 & $ 0.08033 $ 	 & $ 0.00103 $ 	 & $ 0.06699 $ 	 & $ 0.00187 $ 	 & $ 10.35 $ 	 & $ 1.40 $ \\
3573.51203908 	 & $ 0.07965 $ 	 & $ 0.00110 $ 	 & $ 0.06574 $ 	 & $ 0.00209 $ 	 & $ 0.65 $ 	 & $ 1.30 $ \\
3574.52232941 	 & $ 0.07983 $ 	 & $ 0.00109 $ 	 & $ 0.06622 $ 	 & $ 0.00212 $ 	 & $ 9.05 $ 	 & $ 1.10 $ \\
3575.48074894 	 & $ 0.07969 $ 	 & $ 0.00104 $ 	 & $ 0.06625 $ 	 & $ 0.00190 $ 	 & $ 4.35 $ 	 & $ 1.00 $ \\
3576.53604566 	 & $ 0.08035 $ 	 & $ 0.00107 $ 	 & $ 0.06552 $ 	 & $ 0.00205 $ 	 & $ -7.15 $ 	 & $ 1.00 $ \\
3577.59260329 	 & $ 0.08076 $ 	 & $ 0.00110 $ 	 & $ 0.06500 $ 	 & $ 0.00206 $ 	 & $ -10.85 $ 	 & $ 1.20 $ \\
3578.51071262 	 & $ 0.08113 $ 	 & $ 0.00110 $ 	 & $ 0.06370 $ 	 & $ 0.00196 $ 	 & $ 0.35 $ 	 & $ 0.90 $ \\
3578.62960203 	 & $ 0.08012 $ 	 & $ 0.00108 $ 	 & $ 0.06512 $ 	 & $ 0.00198 $ 	 & $ 2.25 $ 	 & $ 1.10 $ \\
3579.46255685 	 & $ 0.08028 $ 	 & $ 0.00108 $ 	 & $ 0.06542 $ 	 & $ 0.00204 $ 	 & $ 13.25 $ 	 & $ 0.90 $ \\
3579.62104900 	 & $ 0.07996 $ 	 & $ 0.00110 $ 	 & $ 0.06424 $ 	 & $ 0.00209 $ 	 & $ 14.75 $ 	 & $ 1.10 $ \\
3585.46177062 	 & $ 0.08093 $ 	 & $ 0.00115 $ 	 & $ 0.06587 $ 	 & $ 0.00213 $ 	 & $ 7.75 $ 	 & $ 1.10 $ \\
3586.46515860 	 & $ 0.08059 $ 	 & $ 0.00111 $ 	 & $ 0.06501 $ 	 & $ 0.00200 $ 	 & $ -3.05 $ 	 & $ 0.80 $ \\
3587.46470413 	 & $ 0.08043 $ 	 & $ 0.00107 $ 	 & $ 0.06608 $ 	 & $ 0.00225 $ 	 & $ -17.25 $ 	 & $ 1.60 $ \\
3588.53806182 	 & $ 0.08028 $ 	 & $ 0.00118 $ 	 & $ 0.06399 $ 	 & $ 0.00250 $ 	 & $ -8.15 $ 	 & $ 2.60 $ \\
3589.46202396 	 & $ 0.08012 $ 	 & $ 0.00109 $ 	 & $ 0.06551 $ 	 & $ 0.00202 $ 	 & $ 6.05 $ 	 & $ 0.80 $ \\
3590.46389515 	 & $ 0.08080 $ 	 & $ 0.00110 $ 	 & $ 0.06395 $ 	 & $ 0.00199 $ 	 & $ 12.75 $ 	 & $ 0.80 $ \\
3591.46648375 	 & $ 0.08112 $ 	 & $ 0.00110 $ 	 & $ 0.06539 $ 	 & $ 0.00206 $ 	 & $ 7.75 $ 	 & $ 0.80 $ \\
3592.46481341 	 & $ 0.08119 $ 	 & $ 0.00108 $ 	 & $ 0.06565 $ 	 & $ 0.00201 $ 	 & $ -5.25 $ 	 & $ 0.80 $ \\
3606.55167944 	 & $ 0.08037 $ 	 & $ 0.00112 $ 	 & $ 0.06543 $ 	 & $ 0.00240 $ 	 & $ 14.85 $ 	 & $ 2.10 $ \\
3607.50752689 	 & $ 0.07987 $ 	 & $ 0.00111 $ 	 & $ 0.06538 $ 	 & $ 0.00197 $ 	 & $ 11.55 $ 	 & $ 1.00 $ \\
3608.48264332 	 & $ 0.08049 $ 	 & $ 0.00112 $ 	 & $ 0.06696 $ 	 & $ 0.00199 $ 	 & $ -3.75 $ 	 & $ 1.20 $ \\
3609.48845185 	 & $ 0.07942 $ 	 & $ 0.00114 $ 	 & $ 0.06465 $ 	 & $ 0.00216 $ 	 & $ -11.35 $ 	 & $ 1.60 $ \\
3757.87731922 	 & $ 0.08183 $ 	 & $ 0.00109 $ 	 & $ 0.06323 $ 	 & $ 0.00205 $ 	 & $ 5.75 $ 	 & $ 1.00 $ \\
3760.87547540 	 & $ 0.08189 $ 	 & $ 0.00108 $ 	 & $ 0.06455 $ 	 & $ 0.00212 $ 	 & $ -1.45 $ 	 & $ 1.30 $ \\
3761.85921591 	 & $ 0.08145 $ 	 & $ 0.00108 $ 	 & $ 0.06180 $ 	 & $ 0.00211 $ 	 & $ 7.45 $ 	 & $ 1.40 $ \\
3811.84694254 	 & $ 0.08018 $ 	 & $ 0.00102 $ 	 & $ 0.06611 $ 	 & $ 0.00218 $ 	 & $ 6.45 $ 	 & $ 1.30 $ \\
3813.82701666 	 & $ 0.07988 $ 	 & $ 0.00105 $ 	 & $ 0.06468 $ 	 & $ 0.00197 $ 	 & $ -9.95 $ 	 & $ 0.90 $ \\
3830.83695707 	 & $ 0.07811 $ 	 & $ 0.00102 $ 	 & $ 0.06759 $ 	 & $ 0.00191 $ 	 & $ -1.75 $ 	 & $ 0.90 $ \\
3862.70144100 	 & $ 0.07917 $ 	 & $ 0.00105 $ 	 & $ 0.06596 $ 	 & $ 0.00200 $ 	 & $ -0.55 $ 	 & $ 0.90 $ \\
3864.71366151 	 & $ 0.08007 $ 	 & $ 0.00105 $ 	 & $ 0.06692 $ 	 & $ 0.00201 $ 	 & $ 13.85 $ 	 & $ 1.10 $ \\
3867.75217074 	 & $ 0.07962 $ 	 & $ 0.00105 $ 	 & $ 0.06653 $ 	 & $ 0.00198 $ 	 & $ -9.25 $ 	 & $ 1.10 $ \\
3870.69660298 	 & $ 0.07954 $ 	 & $ 0.00103 $ 	 & $ 0.06641 $ 	 & $ 0.00202 $ 	 & $ 6.35 $ 	 & $ 1.10 $ \\
3882.65776272 	 & $ 0.08071 $ 	 & $ 0.00104 $ 	 & $ 0.06418 $ 	 & $ 0.00201 $ 	 & $ -11.35 $ 	 & $ 0.90 $ \\
3887.69073769 	 & $ 0.08109 $ 	 & $ 0.00105 $ 	 & $ 0.06347 $ 	 & $ 0.00207 $ 	 & $ -5.65 $ 	 & $ 0.80 $ \\
3918.62175058 	 & $ 0.08368 $ 	 & $ 0.00103 $ 	 & $ 0.06458 $ 	 & $ 0.00207 $ 	 & $ 7.75 $ 	 & $ 1.10 $ \\
3920.59494713 	 & $ 0.08384 $ 	 & $ 0.00106 $ 	 & $ 0.06280 $ 	 & $ 0.00208 $ 	 & $ -18.45 $ 	 & $ 1.00 $ \\
3945.54312171 	 & $ 0.08001 $ 	 & $ 0.00104 $ 	 & $ 0.06621 $ 	 & $ 0.00202 $ 	 & $ 9.25 $ 	 & $ 1.00 $ \\
3951.48592717 	 & $ 0.08063 $ 	 & $ 0.00109 $ 	 & $ 0.06532 $ 	 & $ 0.00201 $ 	 & $ 6.55 $ 	 & $ 0.80 $ \\
3975.47159589 	 & $ 0.08170 $ 	 & $ 0.00106 $ 	 & $ 0.06446 $ 	 & $ 0.00220 $ 	 & $ -7.35 $ 	 & $ 1.00 $ \\
3979.54397490 	 & $ 0.08115 $ 	 & $ 0.00115 $ 	 & $ 0.06458 $ 	 & $ 0.00220 $ 	 & $ -7.35 $ 	 & $ 1.30 $ \\
4166.87418187 	 & $ 0.08226 $ 	 & $ 0.00104 $ 	 & $ 0.06321 $ 	 & $ 0.00205 $ 	 & $ -12.85 $ 	 & $ 1.10 $ \\
4170.85396096 	 & $ 0.08282 $ 	 & $ 0.00107 $ 	 & $ 0.06199 $ 	 & $ 0.00204 $ 	 & $ 7.95 $ 	 & $ 0.90 $ \\
4194.87234931 	 & $ 0.08346 $ 	 & $ 0.00106 $ 	 & $ 0.06611 $ 	 & $ 0.00191 $ 	 & $ -12.45 $ 	 & $ 1.10 $ \\
4196.75038397 	 & $ 0.08160 $ 	 & $ 0.00103 $ 	 & $ 0.06482 $ 	 & $ 0.00213 $ 	 & $ 16.35 $ 	 & $ 1.20 $ \\
4197.84503865 	 & $ 0.08155 $ 	 & $ 0.00106 $ 	 & $ 0.06330 $ 	 & $ 0.00230 $ 	 & $ 15.25 $ 	 & $ 1.20 $ \\
4198.85550900 	 & $ 0.08139 $ 	 & $ 0.00103 $ 	 & $ 0.06301 $ 	 & $ 0.00221 $ 	 & $ -2.15 $ 	 & $ 1.30 $ \\
4199.73286924 	 & $ 0.08138 $ 	 & $ 0.00103 $ 	 & $ 0.06272 $ 	 & $ 0.00207 $ 	 & $ -5.35 $ 	 & $ 1.00 $ \\
4200.91091462 	 & $ 0.08169 $ 	 & $ 0.00101 $ 	 & $ 0.06434 $ 	 & $ 0.00228 $ 	 & $ -0.05 $ 	 & $ 1.10 $ \\
4201.86855390 	 & $ 0.08170 $ 	 & $ 0.00105 $ 	 & $ 0.06385 $ 	 & $ 0.00216 $ 	 & $ 9.35 $ 	 & $ 1.00 $ \\
4202.88259492 	 & $ 0.08150 $ 	 & $ 0.00103 $ 	 & $ 0.06233 $ 	 & $ 0.00206 $ 	 & $ 12.55 $ 	 & $ 1.00 $ \\
4228.74156134 	 & $ 0.08035 $ 	 & $ 0.00101 $ 	 & $ 0.06512 $ 	 & $ 0.00191 $ 	 & $ 8.55 $ 	 & $ 1.10 $ \\
4229.70047777 	 & $ 0.08017 $ 	 & $ 0.00108 $ 	 & $ 0.06406 $ 	 & $ 0.00203 $ 	 & $ 10.35 $ 	 & $ 1.50 $ \\
4230.76213594 	 & $ 0.08038 $ 	 & $ 0.00103 $ 	 & $ 0.06433 $ 	 & $ 0.00196 $ 	 & $ -1.75 $ 	 & $ 1.00 $ \\
4234.64591609 	 & $ 0.08059 $ 	 & $ 0.00105 $ 	 & $ 0.06638 $ 	 & $ 0.00186 $ 	 & $ 14.65 $ 	 & $ 1.20 $ \\
4253.63316819 	 & $ 0.08075 $ 	 & $ 0.00104 $ 	 & $ 0.06581 $ 	 & $ 0.00187 $ 	 & $ -9.35 $ 	 & $ 1.00 $ \\
4254.66480986 	 & $ 0.08029 $ 	 & $ 0.00106 $ 	 & $ 0.06650 $ 	 & $ 0.00189 $ 	 & $ -4.35 $ 	 & $ 1.00 $ \\
4291.56884952 	 & $ 0.08327 $ 	 & $ 0.00102 $ 	 & $ 0.06255 $ 	 & $ 0.00222 $ 	 & $ -6.85 $ 	 & $ 1.40 $ \\
4292.59081443 	 & $ 0.08250 $ 	 & $ 0.00106 $ 	 & $ 0.06281 $ 	 & $ 0.00208 $ 	 & $ 0.25 $ 	 & $ 0.90 $ \\
4293.62586765 	 & $ 0.08206 $ 	 & $ 0.00105 $ 	 & $ 0.06313 $ 	 & $ 0.00213 $ 	 & $ 9.85 $ 	 & $ 1.00 $ \\
4295.63944847 	 & $ 0.08274 $ 	 & $ 0.00110 $ 	 & $ 0.06216 $ 	 & $ 0.00207 $ 	 & $ -10.35 $ 	 & $ 1.10 $ \\
4296.60611381 	 & $ 0.08282 $ 	 & $ 0.00106 $ 	 & $ 0.06367 $ 	 & $ 0.00222 $ 	 & $ -19.65 $ 	 & $ 1.30 $ \\
4297.64193914 	 & $ 0.08281 $ 	 & $ 0.00109 $ 	 & $ 0.06110 $ 	 & $ 0.00205 $ 	 & $ -8.25 $ 	 & $ 1.00 $ \\
4298.56759966 	 & $ 0.08275 $ 	 & $ 0.00109 $ 	 & $ 0.06334 $ 	 & $ 0.00221 $ 	 & $ 7.75 $ 	 & $ 1.10 $ \\
4299.62219505 	 & $ 0.08293 $ 	 & $ 0.00111 $ 	 & $ 0.06222 $ 	 & $ 0.00230 $ 	 & $ 10.95 $ 	 & $ 1.60 $ \\
4300.61910960 	 & $ 0.08277 $ 	 & $ 0.00112 $ 	 & $ 0.06149 $ 	 & $ 0.00206 $ 	 & $ -0.95 $ 	 & $ 1.00 $ \\
4315.50749361 	 & $ 0.08206 $ 	 & $ 0.00110 $ 	 & $ 0.06344 $ 	 & $ 0.00226 $ 	 & $ 14.05 $ 	 & $ 1.70 $ \\
4317.48084703 	 & $ 0.08180 $ 	 & $ 0.00105 $ 	 & $ 0.06215 $ 	 & $ 0.00208 $ 	 & $ -6.95 $ 	 & $ 1.00 $ \\
4319.49052713 	 & $ 0.08156 $ 	 & $ 0.00111 $ 	 & $ 0.05922 $ 	 & $ 0.00228 $ 	 & $ 2.95 $ 	 & $ 1.40 $ \\
4320.54407172 	 & $ 0.08156 $ 	 & $ 0.00105 $ 	 & $ 0.06228 $ 	 & $ 0.00204 $ 	 & $ 10.95 $ 	 & $ 1.00 $ \\
4323.50705017 	 & $ 0.08219 $ 	 & $ 0.00112 $ 	 & $ 0.05650 $ 	 & $ 0.00304 $ 	 & $ -11.05 $ 	 & $ 3.70 $ \\
4340.55578099 	 & $ 0.08191 $ 	 & $ 0.00113 $ 	 & $ 0.06281 $ 	 & $ 0.00210 $ 	 & $ -1.65 $ 	 & $ 0.90 $ \\
4342.48620376 	 & $ 0.08193 $ 	 & $ 0.00111 $ 	 & $ 0.06358 $ 	 & $ 0.00208 $ 	 & $ 19.05 $ 	 & $ 1.10 $ \\
4349.51516265 	 & $ 0.08154 $ 	 & $ 0.00107 $ 	 & $ 0.06443 $ 	 & $ 0.00208 $ 	 & $ -11.85 $ 	 & $ 1.00 $ \\
4530.85566014 	 & $ 0.08240 $ 	 & $ 0.00105 $ 	 & $ 0.06261 $ 	 & $ 0.00203 $ 	 & $ 8.25 $ 	 & $ 1.00 $ \\
4550.83127371 	 & $ 0.08210 $ 	 & $ 0.00101 $ 	 & $ 0.06354 $ 	 & $ 0.00201 $ 	 & $ 6.45 $ 	 & $ 0.90 $ \\
4553.80372171 	 & $ 0.08208 $ 	 & $ 0.00101 $ 	 & $ 0.06325 $ 	 & $ 0.00206 $ 	 & $ -12.85 $ 	 & $ 0.80 $ \\
4563.83800153 	 & $ 0.08207 $ 	 & $ 0.00104 $ 	 & $ 0.06247 $ 	 & $ 0.00209 $ 	 & $ -3.95 $ 	 & $ 0.90 $ \\
4566.76114737 	 & $ 0.08212 $ 	 & $ 0.00105 $ 	 & $ 0.06349 $ 	 & $ 0.00216 $ 	 & $ 0.95 $ 	 & $ 1.10 $ \\
4567.79167115 	 & $ 0.08154 $ 	 & $ 0.00104 $ 	 & $ 0.06256 $ 	 & $ 0.00216 $ 	 & $ 9.35 $ 	 & $ 1.00 $ \\
4569.79330162 	 & $ 0.08218 $ 	 & $ 0.00103 $ 	 & $ 0.06298 $ 	 & $ 0.00208 $ 	 & $ -14.35 $ 	 & $ 1.00 $ \\
4570.80424813 	 & $ 0.08215 $ 	 & $ 0.00105 $ 	 & $ 0.06443 $ 	 & $ 0.00217 $ 	 & $ -14.35 $ 	 & $ 1.00 $ \\
4571.81837609 	 & $ 0.08150 $ 	 & $ 0.00103 $ 	 & $ 0.06390 $ 	 & $ 0.00217 $ 	 & $ 4.15 $ 	 & $ 1.10 $ \\
4587.86196779 	 & $ 0.08238 $ 	 & $ 0.00104 $ 	 & $ 0.06416 $ 	 & $ 0.00213 $ 	 & $ 3.55 $ 	 & $ 1.60 $ \\
4588.83879925 	 & $ 0.08157 $ 	 & $ 0.00107 $ 	 & $ 0.06267 $ 	 & $ 0.00199 $ 	 & $ 9.75 $ 	 & $ 1.10 $ \\
4589.82749271 	 & $ 0.08215 $ 	 & $ 0.00106 $ 	 & $ 0.06274 $ 	 & $ 0.00197 $ 	 & $ 6.05 $ 	 & $ 1.10 $ \\
4590.81963366 	 & $ 0.08232 $ 	 & $ 0.00105 $ 	 & $ 0.06286 $ 	 & $ 0.00198 $ 	 & $ -4.95 $ 	 & $ 1.00 $ \\
4591.81712025 	 & $ 0.08153 $ 	 & $ 0.00107 $ 	 & $ 0.06344 $ 	 & $ 0.00220 $ 	 & $ -19.55 $ 	 & $ 1.60 $ \\
4592.82733676 	 & $ 0.08186 $ 	 & $ 0.00106 $ 	 & $ 0.06280 $ 	 & $ 0.00195 $ 	 & $ -9.75 $ 	 & $ 0.90 $ \\
4610.74293237 	 & $ 0.08769 $ 	 & $ 0.00103 $ 	 & $ 0.07275 $ 	 & $ 0.00157 $ 	 & $ 19.55 $ 	 & $ 1.10 $ \\
4611.71347745 	 & $ 0.08142 $ 	 & $ 0.00107 $ 	 & $ 0.06282 $ 	 & $ 0.00200 $ 	 & $ 9.05 $ 	 & $ 0.90 $ \\
4616.71302788 	 & $ 0.08264 $ 	 & $ 0.00107 $ 	 & $ 0.06292 $ 	 & $ 0.00196 $ 	 & $ 8.75 $ 	 & $ 1.40 $ \\
4639.68650798 	 & $ 0.08244 $ 	 & $ 0.00109 $ 	 & $ 0.06072 $ 	 & $ 0.00202 $ 	 & $ -9.55 $ 	 & $ 1.10 $ \\
4640.65440901 	 & $ 0.08258 $ 	 & $ 0.00105 $ 	 & $ 0.06104 $ 	 & $ 0.00216 $ 	 & $ -10.15 $ 	 & $ 1.40 $ \\
4641.63170643 	 & $ 0.08245 $ 	 & $ 0.00108 $ 	 & $ 0.06149 $ 	 & $ 0.00205 $ 	 & $ 1.65 $ 	 & $ 1.00 $ \\
4643.64499761 	 & $ 0.08232 $ 	 & $ 0.00110 $ 	 & $ 0.06005 $ 	 & $ 0.00223 $ 	 & $ 2.35 $ 	 & $ 1.30 $ \\
4644.58702786 	 & $ 0.08198 $ 	 & $ 0.00102 $ 	 & $ 0.06107 $ 	 & $ 0.00208 $ 	 & $ -9.85 $ 	 & $ 1.30 $ \\
4646.62535674 	 & $ 0.08234 $ 	 & $ 0.00108 $ 	 & $ 0.06091 $ 	 & $ 0.00215 $ 	 & $ -7.05 $ 	 & $ 1.20 $ \\
4647.57911911 	 & $ 0.08227 $ 	 & $ 0.00108 $ 	 & $ 0.05998 $ 	 & $ 0.00208 $ 	 & $ 7.85 $ 	 & $ 1.10 $ \\
4648.48481691 	 & $ 0.08186 $ 	 & $ 0.00108 $ 	 & $ 0.06251 $ 	 & $ 0.00210 $ 	 & $ 10.05 $ 	 & $ 1.10 $ \\
4661.55370583 	 & $ 0.08132 $ 	 & $ 0.00107 $ 	 & $ 0.06268 $ 	 & $ 0.00202 $ 	 & $ -11.65 $ 	 & $ 1.20 $ \\
4662.54940760 	 & $ 0.08056 $ 	 & $ 0.00107 $ 	 & $ 0.06251 $ 	 & $ 0.00209 $ 	 & $ -1.05 $ 	 & $ 1.40 $ \\
4663.54486562 	 & $ 0.08132 $ 	 & $ 0.00106 $ 	 & $ 0.06269 $ 	 & $ 0.00208 $ 	 & $ 14.15 $ 	 & $ 1.20 $ \\
4664.55304174 	 & $ 0.08074 $ 	 & $ 0.00107 $ 	 & $ 0.06230 $ 	 & $ 0.00203 $ 	 & $ 13.25 $ 	 & $ 1.30 $ \\
4665.56937618 	 & $ 0.08124 $ 	 & $ 0.00107 $ 	 & $ 0.06255 $ 	 & $ 0.00203 $ 	 & $ 6.65 $ 	 & $ 1.00 $ \\
4672.53172294 	 & $ 0.08026 $ 	 & $ 0.00109 $ 	 & $ 0.06249 $ 	 & $ 0.00218 $ 	 & $ -4.55 $ 	 & $ 1.90 $ \\
4674.52411958 	 & $ 0.08048 $ 	 & $ 0.00111 $ 	 & $ 0.06394 $ 	 & $ 0.00216 $ 	 & $ 9.35 $ 	 & $ 1.40 $ \\
4677.50511368 	 & $ 0.08072 $ 	 & $ 0.00110 $ 	 & $ 0.06399 $ 	 & $ 0.00207 $ 	 & $ -9.55 $ 	 & $ 1.10 $ \\
4678.55678527 	 & $ 0.08076 $ 	 & $ 0.00113 $ 	 & $ 0.06205 $ 	 & $ 0.00203 $ 	 & $ 2.85 $ 	 & $ 1.20 $ \\
4679.50403353 	 & $ 0.08128 $ 	 & $ 0.00109 $ 	 & $ 0.06462 $ 	 & $ 0.00220 $ 	 & $ 13.25 $ 	 & $ 1.70 $ \\
4681.51414324 	 & $ 0.08134 $ 	 & $ 0.00110 $ 	 & $ 0.06167 $ 	 & $ 0.00215 $ 	 & $ 3.15 $ 	 & $ 1.50 $ \\
4682.50334275 	 & $ 0.08140 $ 	 & $ 0.00107 $ 	 & $ 0.06151 $ 	 & $ 0.00204 $ 	 & $ -5.85 $ 	 & $ 1.40 $ \\
4701.48507398 	 & $ 0.08138 $ 	 & $ 0.00105 $ 	 & $ 0.06400 $ 	 & $ 0.00203 $ 	 & $ 14.15 $ 	 & $ 1.30 $ \\
4703.51304213 	 & $ 0.08203 $ 	 & $ 0.00107 $ 	 & $ 0.06543 $ 	 & $ 0.00212 $ 	 & $ -1.75 $ 	 & $ 1.30 $ \\
4708.47904950 	 & $ 0.08118 $ 	 & $ 0.00106 $ 	 & $ 0.06442 $ 	 & $ 0.00212 $ 	 & $ -5.35 $ 	 & $ 1.20 $ \\
4721.47303256 	 & $ 0.08173 $ 	 & $ 0.00103 $ 	 & $ 0.06714 $ 	 & $ 0.00226 $ 	 & $ -14.45 $ 	 & $ 1.30 $ \\
4722.47237098 	 & $ 0.08166 $ 	 & $ 0.00104 $ 	 & $ 0.06531 $ 	 & $ 0.00225 $ 	 & $ 2.65 $ 	 & $ 1.20 $ \\
4916.91734966 	 & $ 0.08142 $ 	 & $ 0.00105 $ 	 & $ 0.06276 $ 	 & $ 0.00202 $ 	 & $ 4.55 $ 	 & $ 1.00 $ \\
4919.77750721 	 & $ 0.08103 $ 	 & $ 0.00105 $ 	 & $ 0.06266 $ 	 & $ 0.00206 $ 	 & $ -13.95 $ 	 & $ 0.90 $ \\
4935.69136142 	 & $ 0.08138 $ 	 & $ 0.00102 $ 	 & $ 0.06418 $ 	 & $ 0.00213 $ 	 & $ -13.75 $ 	 & $ 1.10 $ \\
4938.77022542 	 & $ 0.08154 $ 	 & $ 0.00108 $ 	 & $ 0.06574 $ 	 & $ 0.00210 $ 	 & $ 4.85 $ 	 & $ 1.10 $ \\
4941.70398890 	 & $ 0.08085 $ 	 & $ 0.00103 $ 	 & $ 0.06532 $ 	 & $ 0.00207 $ 	 & $ -12.45 $ 	 & $ 1.00 $ \\
4946.74298194 	 & $ 0.08089 $ 	 & $ 0.00105 $ 	 & $ 0.06887 $ 	 & $ 0.00205 $ 	 & $ -4.95 $ 	 & $ 1.10 $ \\
4955.79357698 	 & $ 0.08117 $ 	 & $ 0.00109 $ 	 & $ 0.06708 $ 	 & $ 0.00188 $ 	 & $ -6.75 $ 	 & $ 1.00 $ \\
4989.67873633 	 & $ 0.08079 $ 	 & $ 0.00107 $ 	 & $ 0.06836 $ 	 & $ 0.00201 $ 	 & $ -5.55 $ 	 & $ 1.00 $ \\
4993.61155423 	 & $ 0.08180 $ 	 & $ 0.00111 $ 	 & $ 0.06705 $ 	 & $ 0.00194 $ 	 & $ -6.45 $ 	 & $ 1.00 $ \\
5049.51551307 	 & $ 0.08214 $ 	 & $ 0.00108 $ 	 & $ 0.06293 $ 	 & $ 0.00220 $ 	 & $ -0.25 $ 	 & $ 1.30 $ \\
5056.52501198 	 & $ 0.08135 $ 	 & $ 0.00115 $ 	 & $ 0.05881 $ 	 & $ 0.00253 $ 	 & $ 12.45 $ 	 & $ 2.90 $ \\
5227.84095142 	 & $ 0.08066 $ 	 & $ 0.00107 $ 	 & $ 0.06450 $ 	 & $ 0.00199 $ 	 & $ 11.35 $ 	 & $ 1.10 $ \\
5229.88061948 	 & $ 0.07969 $ 	 & $ 0.00108 $ 	 & $ 0.06544 $ 	 & $ 0.00199 $ 	 & $ -5.75 $ 	 & $ 1.40 $ \\
5230.85894510 	 & $ 0.07991 $ 	 & $ 0.00111 $ 	 & $ 0.06717 $ 	 & $ 0.00205 $ 	 & $ -8.15 $ 	 & $ 1.80 $ \\
5232.88302245 	 & $ 0.07982 $ 	 & $ 0.00103 $ 	 & $ 0.06516 $ 	 & $ 0.00197 $ 	 & $ 16.25 $ 	 & $ 1.80 $ \\
5272.83531355 	 & $ 0.08011 $ 	 & $ 0.00103 $ 	 & $ 0.05985 $ 	 & $ 0.00281 $ 	 & $ -2.85 $ 	 & $ 1.10 $ \\
5275.80926058 	 & $ 0.08082 $ 	 & $ 0.00102 $ 	 & $ 0.06117 $ 	 & $ 0.00341 $ 	 & $ 9.25 $ 	 & $ 1.30 $ \\
5277.83040795 	 & $ 0.08134 $ 	 & $ 0.00105 $ 	 & $ 0.06256 $ 	 & $ 0.00192 $ 	 & $ -1.75 $ 	 & $ 1.20 $ \\
5282.86587023 	 & $ 0.08249 $ 	 & $ 0.00106 $ 	 & $ 0.06328 $ 	 & $ 0.00207 $ 	 & $ 1.45 $ 	 & $ 1.20 $ \\
5292.84315218 	 & $ 0.08479 $ 	 & $ 0.00104 $ 	 & $ 0.06355 $ 	 & $ 0.00204 $ 	 & $ 7.45 $ 	 & $ 1.10 $ \\
5294.77401958 	 & $ 0.08422 $ 	 & $ 0.00101 $ 	 & $ 0.06239 $ 	 & $ 0.00209 $ 	 & $ -17.25 $ 	 & $ 1.00 $ \\
5295.68471470 	 & $ 0.08380 $ 	 & $ 0.00104 $ 	 & $ 0.06330 $ 	 & $ 0.00215 $ 	 & $ -13.45 $ 	 & $ 1.30 $ \\
5297.76796578 	 & $ 0.08419 $ 	 & $ 0.00103 $ 	 & $ 0.06233 $ 	 & $ 0.00214 $ 	 & $ 10.05 $ 	 & $ 1.00 $ \\
5298.73451600 	 & $ 0.08492 $ 	 & $ 0.00105 $ 	 & $ 0.06320 $ 	 & $ 0.00213 $ 	 & $ 4.05 $ 	 & $ 1.00 $ \\
5299.68211722 	 & $ 0.08523 $ 	 & $ 0.00108 $ 	 & $ 0.06602 $ 	 & $ 0.00226 $ 	 & $ -7.15 $ 	 & $ 1.70 $ \\
5300.72869129 	 & $ 0.08378 $ 	 & $ 0.00108 $ 	 & $ 0.06077 $ 	 & $ 0.00210 $ 	 & $ -15.15 $ 	 & $ 1.30 $ \\
5301.84322951 	 & $ 0.08335 $ 	 & $ 0.00105 $ 	 & $ 0.06259 $ 	 & $ 0.00213 $ 	 & $ -6.25 $ 	 & $ 1.20 $ \\
5305.80849866 	 & $ 0.08292 $ 	 & $ 0.00104 $ 	 & $ 0.06334 $ 	 & $ 0.00210 $ 	 & $ -12.65 $ 	 & $ 1.00 $ \\
5306.76723577 	 & $ 0.08336 $ 	 & $ 0.00103 $ 	 & $ 0.06313 $ 	 & $ 0.00210 $ 	 & $ -6.85 $ 	 & $ 1.00 $ \\
5307.76067164 	 & $ 0.08230 $ 	 & $ 0.00105 $ 	 & $ 0.06295 $ 	 & $ 0.00206 $ 	 & $ 8.65 $ 	 & $ 0.80 $ \\
5308.75780976 	 & $ 0.08244 $ 	 & $ 0.00104 $ 	 & $ 0.06196 $ 	 & $ 0.00199 $ 	 & $ 17.75 $ 	 & $ 0.90 $ \\
5309.76544430 	 & $ 0.08182 $ 	 & $ 0.00104 $ 	 & $ 0.06430 $ 	 & $ 0.00207 $ 	 & $ 4.45 $ 	 & $ 0.90 $ \\
5321.70851551 	 & $ 0.08024 $ 	 & $ 0.00104 $ 	 & $ 0.06460 $ 	 & $ 0.00216 $ 	 & $ -5.65 $ 	 & $ 1.00 $ \\
5325.66237320 	 & $ 0.07960 $ 	 & $ 0.00105 $ 	 & $ 0.06479 $ 	 & $ 0.00205 $ 	 & $ 5.45 $ 	 & $ 1.30 $ \\
5326.61457483 	 & $ 0.07948 $ 	 & $ 0.00105 $ 	 & $ 0.06378 $ 	 & $ 0.00208 $ 	 & $ -9.55 $ 	 & $ 1.30 $ \\
5328.63743408 	 & $ 0.07958 $ 	 & $ 0.00101 $ 	 & $ 0.06403 $ 	 & $ 0.00225 $ 	 & $ -3.65 $ 	 & $ 1.60 $ \\
5334.66358749 	 & $ 0.08001 $ 	 & $ 0.00102 $ 	 & $ 0.06579 $ 	 & $ 0.00214 $ 	 & $ 14.05 $ 	 & $ 1.30 $ \\
5336.78989062 	 & $ 0.07974 $ 	 & $ 0.00106 $ 	 & $ 0.06736 $ 	 & $ 0.00195 $ 	 & $ 5.05 $ 	 & $ 1.00 $ \\
5337.65472672 	 & $ 0.08076 $ 	 & $ 0.00104 $ 	 & $ 0.06753 $ 	 & $ 0.00205 $ 	 & $ -7.25 $ 	 & $ 1.40 $ \\
5349.63633994 	 & $ 0.08021 $ 	 & $ 0.00108 $ 	 & $ 0.06629 $ 	 & $ 0.00220 $ 	 & $ 3.85 $ 	 & $ 3.00 $ \\
5353.57755608 	 & $ 0.07966 $ 	 & $ 0.00106 $ 	 & $ 0.06567 $ 	 & $ 0.00186 $ 	 & $ -11.25 $ 	 & $ 1.30 $ \\
5354.60680719 	 & $ 0.08016 $ 	 & $ 0.00103 $ 	 & $ 0.06506 $ 	 & $ 0.00193 $ 	 & $ -16.35 $ 	 & $ 1.30 $ \\
5355.53696386 	 & $ 0.08052 $ 	 & $ 0.00107 $ 	 & $ 0.06516 $ 	 & $ 0.00188 $ 	 & $ -1.15 $ 	 & $ 1.30 $ \\
5359.56247042 	 & $ 0.08126 $ 	 & $ 0.00107 $ 	 & $ 0.06532 $ 	 & $ 0.00192 $ 	 & $ -7.75 $ 	 & $ 1.20 $ \\
5370.57817460 	 & $ 0.07979 $ 	 & $ 0.00107 $ 	 & $ 0.05973 $ 	 & $ 0.00367 $ 	 & $ -11.45 $ 	 & $ 2.00 $ \\
5372.55365945 	 & $ 0.08066 $ 	 & $ 0.00114 $ 	 & $ 0.06559 $ 	 & $ 0.00213 $ 	 & $ 12.65 $ 	 & $ 1.50 $ \\
5373.60233934 	 & $ 0.08073 $ 	 & $ 0.00109 $ 	 & $ 0.06293 $ 	 & $ 0.00203 $ 	 & $ 11.75 $ 	 & $ 1.20 $ \\
5374.61617082 	 & $ 0.08046 $ 	 & $ 0.00105 $ 	 & $ 0.06330 $ 	 & $ 0.00205 $ 	 & $ -2.35 $ 	 & $ 1.50 $ \\
5375.55662662 	 & $ 0.08059 $ 	 & $ 0.00106 $ 	 & $ 0.06410 $ 	 & $ 0.00205 $ 	 & $ -9.45 $ 	 & $ 1.40 $ \\
5389.64755703 	 & $ 0.08043 $ 	 & $ 0.00110 $ 	 & $ 0.06209 $ 	 & $ 0.00202 $ 	 & $ 11.95 $ 	 & $ 1.50 $ \\
5390.54432217 	 & $ 0.08028 $ 	 & $ 0.00113 $ 	 & $ 0.04905 $ 	 & $ 0.00302 $ 	 & $ 4.65 $ 	 & $ 4.70 $ \\
5391.54670272 	 & $ 0.08079 $ 	 & $ 0.00103 $ 	 & $ 0.06603 $ 	 & $ 0.00196 $ 	 & $ -13.25 $ 	 & $ 1.40 $ \\
5396.49708365 	 & $ 0.08000 $ 	 & $ 0.00106 $ 	 & $ 0.06604 $ 	 & $ 0.00229 $ 	 & $ -8.45 $ 	 & $ 2.10 $ \\
5399.54016828 	 & $ 0.08106 $ 	 & $ 0.00107 $ 	 & $ 0.06579 $ 	 & $ 0.00188 $ 	 & $ 18.65 $ 	 & $ 1.30 $ \\
5401.52230318 	 & $ 0.08054 $ 	 & $ 0.00109 $ 	 & $ 0.06352 $ 	 & $ 0.00227 $ 	 & $ -0.25 $ 	 & $ 1.70 $ \\
5407.49699247 	 & $ 0.08095 $ 	 & $ 0.00106 $ 	 & $ 0.06430 $ 	 & $ 0.00194 $ 	 & $ -11.35 $ 	 & $ 1.00 $ \\
5408.50167800 	 & $ 0.08073 $ 	 & $ 0.00103 $ 	 & $ 0.06331 $ 	 & $ 0.00221 $ 	 & $ -5.05 $ 	 & $ 1.90 $ \\
5410.55603221 	 & $ 0.08142 $ 	 & $ 0.00113 $ 	 & $ 0.06728 $ 	 & $ 0.00188 $ 	 & $ 16.85 $ 	 & $ 1.20 $ \\
5411.51484151 	 & $ 0.08386 $ 	 & $ 0.00108 $ 	 & $ 0.06561 $ 	 & $ 0.00195 $ 	 & $ 8.75 $ 	 & $ 1.50 $ \\
5423.51170893 	 & $ 0.08197 $ 	 & $ 0.00108 $ 	 & $ 0.06384 $ 	 & $ 0.00207 $ 	 & $ -11.05 $ 	 & $ 1.10 $ \\
5427.49845525 	 & $ 0.08560 $ 	 & $ 0.00108 $ 	 & $ 0.06499 $ 	 & $ 0.00194 $ 	 & $ 8.55 $ 	 & $ 1.10 $ \\
5428.48093062 	 & $ 0.08305 $ 	 & $ 0.00106 $ 	 & $ 0.06267 $ 	 & $ 0.00215 $ 	 & $ -4.75 $ 	 & $ 1.30 $ \\
5434.51127301 	 & $ 0.08306 $ 	 & $ 0.00106 $ 	 & $ 0.06361 $ 	 & $ 0.00205 $ 	 & $ -15.65 $ 	 & $ 1.20 $ \\
5435.48704963 	 & $ 0.08317 $ 	 & $ 0.00107 $ 	 & $ 0.06207 $ 	 & $ 0.00208 $ 	 & $ -11.65 $ 	 & $ 1.00 $ \\
5436.48340340 	 & $ 0.08284 $ 	 & $ 0.00106 $ 	 & $ 0.06282 $ 	 & $ 0.00214 $ 	 & $ 3.65 $ 	 & $ 0.90 $ \\
5437.51431457 	 & $ 0.08276 $ 	 & $ 0.00109 $ 	 & $ 0.06251 $ 	 & $ 0.00209 $ 	 & $ 14.85 $ 	 & $ 0.90 $ \\
5439.48708402 	 & $ 0.08228 $ 	 & $ 0.00109 $ 	 & $ 0.06168 $ 	 & $ 0.00215 $ 	 & $ -9.55 $ 	 & $ 1.00 $ \\
5443.49985669 	 & $ 0.08275 $ 	 & $ 0.00115 $ 	 & $ 0.06197 $ 	 & $ 0.00239 $ 	 & $ 2.95 $ 	 & $ 2.10 $ \\
5444.48950218 	 & $ 0.08236 $ 	 & $ 0.00108 $ 	 & $ 0.05653 $ 	 & $ 0.00305 $ 	 & $ -10.55 $ 	 & $ 1.10 $ \\
5445.49327897 	 & $ 0.08190 $ 	 & $ 0.00108 $ 	 & $ 0.06341 $ 	 & $ 0.00224 $ 	 & $ -21.95 $ 	 & $ 1.20 $ \\
5450.48002349 	 & $ 0.08130 $ 	 & $ 0.00108 $ 	 & $ 0.06307 $ 	 & $ 0.00216 $ 	 & $ -10.35 $ 	 & $ 1.40 $ \\
5453.48660148 	 & $ 0.08145 $ 	 & $ 0.00112 $ 	 & $ 0.06394 $ 	 & $ 0.00219 $ 	 & $ 14.85 $ 	 & $ 1.20 $ \\
5454.47679838 	 & $ 0.08125 $ 	 & $ 0.00105 $ 	 & $ 0.06584 $ 	 & $ 0.00238 $ 	 & $ 4.15 $ 	 & $ 1.40 $ \\
5455.48896200 	 & $ 0.08098 $ 	 & $ 0.00109 $ 	 & $ 0.06401 $ 	 & $ 0.00214 $ 	 & $ -9.45 $ 	 & $ 1.30 $ \\
5457.47396646 	 & $ 0.08089 $ 	 & $ 0.00106 $ 	 & $ 0.06624 $ 	 & $ 0.00201 $ 	 & $ -4.35 $ 	 & $ 1.00 $ \\
5458.48996366 	 & $ 0.08064 $ 	 & $ 0.00108 $ 	 & $ 0.06546 $ 	 & $ 0.00209 $ 	 & $ 6.65 $ 	 & $ 1.30 $ \\
5464.48161306 	 & $ 0.08033 $ 	 & $ 0.00107 $ 	 & $ 0.06409 $ 	 & $ 0.00222 $ 	 & $ 16.45 $ 	 & $ 1.70 $ \\
5626.90847403 	 & $ 0.07998 $ 	 & $ 0.00107 $ 	 & $ 0.06121 $ 	 & $ 0.00208 $ 	 & $ -3.85 $ 	 & $ 1.70 $ \\
5627.86993694 	 & $ 0.08031 $ 	 & $ 0.00108 $ 	 & $ 0.06323 $ 	 & $ 0.00210 $ 	 & $ -16.65 $ 	 & $ 1.20 $ \\
5629.88249802 	 & $ 0.08046 $ 	 & $ 0.00105 $ 	 & $ 0.06304 $ 	 & $ 0.00194 $ 	 & $ 7.15 $ 	 & $ 1.00 $ \\
5630.88944867 	 & $ 0.08065 $ 	 & $ 0.00106 $ 	 & $ 0.06230 $ 	 & $ 0.00209 $ 	 & $ 14.55 $ 	 & $ 1.20 $ \\
5633.83855319 	 & $ 0.07954 $ 	 & $ 0.00105 $ 	 & $ 0.06280 $ 	 & $ 0.00197 $ 	 & $ -11.05 $ 	 & $ 0.90 $ \\
5634.83780289 	 & $ 0.08131 $ 	 & $ 0.00105 $ 	 & $ 0.06303 $ 	 & $ 0.00202 $ 	 & $ 1.45 $ 	 & $ 1.10 $ \\
5635.80037040 	 & $ 0.08145 $ 	 & $ 0.00104 $ 	 & $ 0.06229 $ 	 & $ 0.00205 $ 	 & $ 14.95 $ 	 & $ 1.40 $ \\
5638.87580215 	 & $ 0.08186 $ 	 & $ 0.00105 $ 	 & $ 0.06353 $ 	 & $ 0.00202 $ 	 & $ -13.95 $ 	 & $ 1.10 $ \\
5639.82564487 	 & $ 0.08131 $ 	 & $ 0.00105 $ 	 & $ 0.06120 $ 	 & $ 0.00201 $ 	 & $ -8.45 $ 	 & $ 1.00 $ \\
5641.85816153 	 & $ 0.08086 $ 	 & $ 0.00103 $ 	 & $ 0.06407 $ 	 & $ 0.00200 $ 	 & $ 9.75 $ 	 & $ 1.00 $ \\
5642.78864791 	 & $ 0.08168 $ 	 & $ 0.00105 $ 	 & $ 0.06322 $ 	 & $ 0.00208 $ 	 & $ -0.75 $ 	 & $ 1.20 $ \\
5644.87267782 	 & $ 0.08130 $ 	 & $ 0.00105 $ 	 & $ 0.06306 $ 	 & $ 0.00206 $ 	 & $ -5.45 $ 	 & $ 1.10 $ \\
5646.85118609 	 & $ 0.08075 $ 	 & $ 0.00105 $ 	 & $ 0.06294 $ 	 & $ 0.00209 $ 	 & $ 12.95 $ 	 & $ 1.20 $ \\
5647.86059610 	 & $ 0.08116 $ 	 & $ 0.00103 $ 	 & $ 0.06037 $ 	 & $ 0.00206 $ 	 & $ 1.85 $ 	 & $ 1.00 $ \\
5648.89759932 	 & $ 0.08162 $ 	 & $ 0.00101 $ 	 & $ 0.06159 $ 	 & $ 0.00198 $ 	 & $ -9.95 $ 	 & $ 1.10 $ \\
5652.83977562 	 & $ 0.08195 $ 	 & $ 0.00105 $ 	 & $ 0.06204 $ 	 & $ 0.00200 $ 	 & $ 3.75 $ 	 & $ 1.10 $ \\
5653.72223734 	 & $ 0.08276 $ 	 & $ 0.00104 $ 	 & $ 0.06077 $ 	 & $ 0.00195 $ 	 & $ -8.65 $ 	 & $ 1.10 $ \\
5654.68243413 	 & $ 0.08150 $ 	 & $ 0.00104 $ 	 & $ 0.06293 $ 	 & $ 0.00194 $ 	 & $ -15.85 $ 	 & $ 1.10 $ \\
5656.75878190 	 & $ 0.08220 $ 	 & $ 0.00103 $ 	 & $ 0.06070 $ 	 & $ 0.00205 $ 	 & $ 6.15 $ 	 & $ 1.00 $ \\
5657.76630370 	 & $ 0.08153 $ 	 & $ 0.00103 $ 	 & $ 0.06208 $ 	 & $ 0.00206 $ 	 & $ 14.25 $ 	 & $ 1.10 $ \\
5658.82034278 	 & $ 0.08213 $ 	 & $ 0.00104 $ 	 & $ 0.06020 $ 	 & $ 0.00214 $ 	 & $ 4.65 $ 	 & $ 1.50 $ \\
5662.76574096 	 & $ 0.08131 $ 	 & $ 0.00105 $ 	 & $ 0.06225 $ 	 & $ 0.00224 $ 	 & $ 14.55 $ 	 & $ 1.40 $ \\
5663.75875103 	 & $ 0.08150 $ 	 & $ 0.00106 $ 	 & $ 0.05926 $ 	 & $ 0.00294 $ 	 & $ 7.25 $ 	 & $ 3.00 $ \\
5672.71848380 	 & $ 0.08133 $ 	 & $ 0.00103 $ 	 & $ 0.06147 $ 	 & $ 0.00205 $ 	 & $ 8.25 $ 	 & $ 1.20 $ \\
5674.73186887 	 & $ 0.08108 $ 	 & $ 0.00103 $ 	 & $ 0.06185 $ 	 & $ 0.00204 $ 	 & $ 3.35 $ 	 & $ 1.00 $ \\
5675.77838073 	 & $ 0.08134 $ 	 & $ 0.00109 $ 	 & $ 0.06366 $ 	 & $ 0.00214 $ 	 & $ -12.15 $ 	 & $ 1.30 $ \\
5676.75948402 	 & $ 0.08059 $ 	 & $ 0.00106 $ 	 & $ 0.06384 $ 	 & $ 0.00212 $ 	 & $ -9.05 $ 	 & $ 1.40 $ \\
5677.69188083 	 & $ 0.08090 $ 	 & $ 0.00106 $ 	 & $ 0.06324 $ 	 & $ 0.00228 $ 	 & $ 3.75 $ 	 & $ 1.50 $ \\
5678.76651339 	 & $ 0.08143 $ 	 & $ 0.00109 $ 	 & $ 0.06323 $ 	 & $ 0.00216 $ 	 & $ 11.25 $ 	 & $ 1.50 $ \\
5679.71364130 	 & $ 0.08148 $ 	 & $ 0.00105 $ 	 & $ 0.06345 $ 	 & $ 0.00206 $ 	 & $ 6.45 $ 	 & $ 0.90 $ \\
5680.62401932 	 & $ 0.08091 $ 	 & $ 0.00108 $ 	 & $ 0.06332 $ 	 & $ 0.00210 $ 	 & $ -2.15 $ 	 & $ 1.30 $ \\
5681.67630837 	 & $ 0.08351 $ 	 & $ 0.00105 $ 	 & $ 0.06415 $ 	 & $ 0.00196 $ 	 & $ -12.95 $ 	 & $ 1.00 $ \\
5682.67266874 	 & $ 0.08165 $ 	 & $ 0.00107 $ 	 & $ 0.06284 $ 	 & $ 0.00209 $ 	 & $ -2.25 $ 	 & $ 1.00 $ \\
5683.62142257 	 & $ 0.08130 $ 	 & $ 0.00104 $ 	 & $ 0.06412 $ 	 & $ 0.00205 $ 	 & $ 13.45 $ 	 & $ 1.00 $ \\
5684.67392635 	 & $ 0.08185 $ 	 & $ 0.00104 $ 	 & $ 0.06160 $ 	 & $ 0.00199 $ 	 & $ 14.35 $ 	 & $ 1.10 $ \\
5685.64645049 	 & $ 0.08160 $ 	 & $ 0.00105 $ 	 & $ 0.06368 $ 	 & $ 0.00205 $ 	 & $ 1.55 $ 	 & $ 0.90 $ \\
5686.65353361 	 & $ 0.08118 $ 	 & $ 0.00104 $ 	 & $ 0.06292 $ 	 & $ 0.00217 $ 	 & $ -7.25 $ 	 & $ 1.40 $ \\
5689.71145134 	 & $ 0.08181 $ 	 & $ 0.00105 $ 	 & $ 0.05293 $ 	 & $ 0.00430 $ 	 & $ 12.55 $ 	 & $ 1.30 $ \\
5690.73996295 	 & $ 0.08166 $ 	 & $ 0.00108 $ 	 & $ 0.06031 $ 	 & $ 0.00204 $ 	 & $ -1.15 $ 	 & $ 1.60 $ \\
5691.69250044 	 & $ 0.08241 $ 	 & $ 0.00105 $ 	 & $ 0.06247 $ 	 & $ 0.00209 $ 	 & $ -11.95 $ 	 & $ 1.40 $ \\
5692.71193460 	 & $ 0.08201 $ 	 & $ 0.00108 $ 	 & $ 0.06362 $ 	 & $ 0.00202 $ 	 & $ -12.75 $ 	 & $ 1.20 $ \\
5693.75657536 	 & $ 0.08151 $ 	 & $ 0.00105 $ 	 & $ 0.06438 $ 	 & $ 0.00196 $ 	 & $ -2.25 $ 	 & $ 1.00 $ \\
5695.62766961 	 & $ 0.08210 $ 	 & $ 0.00104 $ 	 & $ 0.06189 $ 	 & $ 0.00203 $ 	 & $ 12.15 $ 	 & $ 0.90 $ \\
5711.55591366 	 & $ 0.08252 $ 	 & $ 0.00123 $ 	 & $ 0.06046 $ 	 & $ 0.00468 $ 	 & $ \cdots $ 	 & $ \cdots $ \\
5749.54860147 	 & $ 0.08002 $ 	 & $ 0.00104 $ 	 & $ 0.06499 $ 	 & $ 0.00209 $ 	 & $ \cdots $ 	 & $ \cdots $ \\
5754.63524961 	 & $ 0.07902 $ 	 & $ 0.00106 $ 	 & $ 0.07108 $ 	 & $ 0.00221 $ 	 & $ \cdots $ 	 & $ \cdots $ \\
\end{supertabular}
\end{center}


\end{document}